\documentclass[12pt,a4paper]{article}
\usepackage{graphicx}
\usepackage{jcappub}
\usepackage{epsfig} 
\usepackage{amsmath}
\usepackage{amsfonts}
\usepackage{graphicx}
\usepackage{amssymb}

\title{The Non-Gaussian Halo Mass Function and Non-Spherical Halo Collapse: Theory vs. Simulations}
\author[*]{Ixandra E. Achitouv}
\author[*]{$\&$ Pier Stefano Corasaniti}
\affiliation[*]{Laboratoire Univers et Th\'eories (LUTh),\\ UMR 8102
CNRS, Observatoire de Paris, Universit\'e Paris Diderot, \\ 5 Place
  Jules Janssen, 92190 Meudon, France}
\emailAdd{Ixandra.Achitouv@obspm.fr, Pier-Stefano.Corasaniti@obspm.fr}

\abstract{ 

The mass distribution of dark matter halos  
is a sensitive probe of primordial non-Gaussianity (NG). We derive 
an analytical formula of the halo mass function by 
perturbatively computing the excursion set path-integrals for a 
non-Gaussian density field with non-vanishing skewness, $f_{NL}$. 
We assume a stochastic barrier model which captures the main  
features of the ellipsoidal collapse of halos.  
Contrary to previous results based on extensions of the 
Press-Schechter formalism to NG initial conditions,  
we find that the non-spherical collapse of halos directly alter  
the signature of primordial NG.  
This points toward a potential degeneracy between the effect 
of primordial non-Gaussianity and that of non-linear halo collapse.  
The inferred mass function is found to be in remarkable agreement with 
N-body simulations of NG local type. Deviations are well within numerical 
uncertainties for all values of $-80<f_{NL}^{\rm loc}<300$ in the range of validity of 
the perturbative calculation.  
Moreover, the comparison with simulation results suggests  
that for $f_{NL}>150 $ or $f_{NL}<-50$ the non-linear collapse of halos, as described by our barrier 
model, strongly deviates from that of Gaussian initial conditions. This is 
not surprising since the effect of non-linear 
gravitational processes may be altered by initially large NG.  
Hence, in the lack of prior theoretical knowledge, 
halo collapse model parameters should be included 
in statistical halo mass function data analysis which 
aim to constrain the signature of primordial NG.}
\begin{document}
\maketitle

\section{Introduction}
The statistics of the primordial density fluctuation field
carries unique information on the physics of the very early Universe. 
Standard slow-roll inflation predicts a nearly scale
invariant spectrum of Gaussian, adiabatic density perturbations
\cite{GuthPi1982,Starobinsky1982,Bardeenetal1983,Falketal1993,
Ganguietal1994,Acquavivaetal2003,Maldacena2003}. 
Alternatively, high-energy physics inspired scenarios may leave
large non-Gaussian signatures on the initial matter density
distribution \cite{ArkaniHamedeetal2004,
Lythetal2003,Creminelli2003,Tong2004,Senatore2005,Chenetal2006,
Bernardeau2007,Barnaby2007,Koyamaetal2007,Holman2008,Khoury2009,Shiu2011}. 
Hence, the detection of primordial non-Gaussianity (NG) has the
potential to disclose quantum physical processes at energies which are
out of reach of laboratory experiments. In recent years this has motivated a revived
interest on testing the Gaussian hypothesis
through measurements of the high-order angular correlation functions of the 
Cosmic Microwave Background (CMB) anisotropies
\cite{Komatsu2003,Creminellietal06,Creminellietal07,Yadav08}. 

The quest for primordial NG
has primarily focused on the amplitude and shape of the
CMB bispectrum statistics
\cite{Babic2004,Creminelli07,Fergusson08,Meerburg09}.
The next generation of cosmic structure surveys may also
provide complementary constraints, since the spatial distribution of
galaxies in the universe carries an imprint of the statistics of the
initial matter density fluctuations.
As shown in several studies, measurements of the galaxy 
bispetrum \cite{Fry1994,Durrer2000,Feldman2001,Scoccimarro2004,Sefusatti2007} 
and the abundance of galaxy clusters 
\cite{LucchinMatarrese1988,Colafrancesco1989,Chui1998} are sensitive
NG probes. However, an accurate modeling of the non-linear
regime of gravitational collapse is
necessary for such tests to be successful. This is because the late 
time non-linear evolution of density perturbations 
induces strong non-Gaussian features in the
spatial distribution of cosmic structures, thus potentially
obscuring the imprint of primordial non-Gaussianity. 
Accounting for these effects is the premise to testing
the statistics of the initial density field.

In this context the determination of the mass distribution of dark
matter halos is of primary importance. Halos are gravitationally bound
objects resulting from the non-linear collapse of dark matter
perturbations. It is inside these virialized structures 
that cooling baryonic gas falls in to form the stars and galaxies
that we observe today. Numerical N-body simulations have been extensively used to study the 
non-linear clustering of matter in the standard cosmological scenario
with Gaussian initial conditions. On the other hand, simulations with non-Gaussian initial
conditions have become available only in the past few years
(e.g., see refs.~\cite{Desjacques09,Grossi09,Pillepich10,Wagner10}). 
This has driven a major effort toward formulating a
mathematical model description of the NG halo mass distribution. 
A number of studies have approached this problem by extending the original Press-Schechther (PS)
derivation to NG initial conditions. The seminal work by Press and Schechter \cite{PressSchechter1974} 
has provided the first formal derivation of the halo abundance
in the case of an initial Gaussian density field. The basic idea is that halos form in regions 
of the linear density field smoothed on a given scale $R$ 
(associated to a mass $M$), whose density lies above a critical 
threshold of collapse (e.g. $\delta_c$ the linearly extrapolated spherical
top-hat density perturbation at the time of collapse \cite{GunnGott1972}). 
Using such a prescription, the fraction of mass in halos with mass $>M$ is given by
\begin{equation}
F(>M)=\int_{\delta_c}^\infty \Pi(\delta,M)d\delta,\label{frac}
\end{equation}
where $\Pi(\delta,M)$ is the probability
distribution of the smoothed linear density field. Then, the 
number density of halos with mass in the range $[M,M+dM]$ is obtained
by equating
\begin{equation}
\frac{dn}{dM}=\frac{\bar{\rho}}{M}\frac{dF}{dM},\label{mfm}
\end{equation}
with $\bar{\rho}$ being the mean background matter density.

The computation can be extended to non-Gaussian initial conditions by
specifying the form of $\Pi(\delta,M)$. For example, Matarrese, Verde and Jimenez \cite{Matarrese2000} have
written $\Pi(\delta,M)$ as a path-integral 
of an infinite series of cumulants of the density field $\delta$. 
This allowed them to derive an approximate formula for $dn/dM$ in
the limit of small NG up to leading order in the skewness parameter, $f_{NL}$.
Similarly, in \cite{Loverdeetal2008} the authors have 
approximated $\Pi(\delta,M)$ as an Edgeworth expansion in the
cumulants and truncated the series to infer an expression of the mass
function which explicitely depends on $f_{NL}$ 
(for an extension to the ellipsoidal collapse threshold see
ref.~\cite{LamSheth2009}).
Overall these studies have provided indications of how
primordial NG affects the halo mass distribution. However, by relying
on the PS approach they have failed to address a number of important
issues that are the main limitation of the PS formalism. As an example in Press-Schechter 
the normalization factor of the mass function 
is related to the ``cloud-in-cloud'' problem. 
Namely, the fact that a collapsed 
region of mass $M$ contributes to the halo mass counting 
only if it is not embedded in a collapsed region of greater mass
$M'>M$. The PS formalism does not address this issue, thus miscounting
the fraction of mass in halos. Moreover, even including an ad-hoc 
correction to recover the correct normalization, 
the PS formula still fails to reproduce results from
N-body simulations. Because of this, analyses based on the PS
approach have mainly focused on predicting the ratio of the
non-Gaussian to Gaussian Press-Schechter mass function
and normalized the result with a fitting function to 
Gaussian N-body simulation data. A recent comparison with non-Gaussian simulations
indicates that these analytical formulae are in relatively good agreement 
with numerical results provided an ad-hoc rescaling of the spherical
collapse threshold is assumed \cite{Grossi09}. 

Ideally, one would like to perform an {\it ab initio}
calculation to consistently account for the cloud-in-cloud problem as
well as the non-spherical collapse of halos for any type of initial
conditions. The excursion set theory introduced by Bond et
al. \cite{Bond1991} provides a powerful mathematical framework capable
of addressing these issues and allowing for a self-consistent derivation of 
the mass function from a limited set of initial assumptions. 
The formalism frames the original Press-Schechter idea in the context 
of stochastic calculus. However, the difficulty in obtaining
analytical solutions for realistic description of halos has been the
main limitation of the theory (for a review see
ref.~\cite{Zentner2007}). 
In fact, in the case of a Gaussian field an analytic treatment is
possible only if one assume an unphysical halo mass definition.
The computation for a realistic case
requires to solve numerically the stochastic model equations
through time consuming Monte Carlo
simulations (see e.g. refs.~\cite{Bond1991,Percival2001}).

In a series of papers Maggiore \& Riotto 
\cite{MaggioreRiotto2010a,MaggioreRiotto2010b,MaggioreRiotto2010c} 
have introduced a path-integral description of the excursion set 
which allows for a perturbative calculation of the 
mass function in the case of physical mass definitions. 
Non-Gaussian initial conditions can be easily implemented in this
novel formulation as well as more realistic conditions for the 
gravitational collapse of halos. This has provided a major breakthrough 
towards an accurate physical model description of the mass function.
For instance, in \cite{PSCIA2011a,PSCIA2011b} we have implemented such an approach 
with an effective stochastic barrier model which accounts for the main 
characteristics of the ellipsoidal collapse of halos. This has allowed
us to derived an analytic formula which shows
an unprecedented agreement with the mass function 
from Gaussian N-body simulations. 

In this paper we extend the study of the non-spherical collapse mass
function to the case of primordial non-Gaussianity. Our goal is to
derive a better understanding of how the halo collapse alters the
signature of primordial NG and confront the theoretical model
prediction against non-Gaussian N-body simulation results. The paper is
organized as follows: in section~\ref{excpath} we introduce the
excursion set theory, the path-integral formalism and discuss
diffusive drifting barrier model of halo collapse. In section~\ref{ngmass}
we derive the non-Gaussian mass function and shown its application
to local and equilateral non-Gaussianity in section~\ref{locequi}. 
We discuss the imprints on the halo mass function due to primordial NG and the halo collapse
model in section~\ref{imprints}, while in section~\ref{nbodycomparison} we present the results of the
comparison to non-Gaussian N-body simulations. Finally we discuss the
conclusions in section~\ref{conclu}.

\section{Excursion Set Theory}\label{excpath}
\subsection{Correlated Random Walks and Halo Mass}
Let us consider a non-Gaussian density field $\delta(\textbf{x})$ smoothed on a
scale $R$:
\begin{equation}
\delta(\textbf{x},R)=\frac{1}{(2\pi)^3}\int d^3k\,\tilde{W}(k,R)\tilde{\delta}(\textbf{k}) e^{-i\textbf{k x}}
\end{equation}
where $\tilde{W}(k,R)$ is the Fourier transform of the
filter function in real space, $W(x,R)$. The latter defines the mass
inside a region of radius $R$ as $M=\bar{\rho}V(R)$, with
$\bar{\rho}$ being the mean background density and $V(R)=\int d^3x\,
W(x,R)$ is the filtered volume. 
Then, the variance of the smoothed density field reads as 
\begin{equation}
S=\langle\delta^2(R)\rangle\equiv\sigma^2(R[M])=\frac{1}{2\pi^2}\int dk~k^2P(k)\tilde{W}^2[k,R(M)],
\end{equation}
where $P(k)$ is the linear matter power spectrum at redshift
$z=0$. Given the one-to-one relation between smoothing radius, mass
and variance, we can rewrite the halo mass function eq.~(\ref{mfm}) as
\begin{equation}
\frac{dn}{dM}=f(\sigma)\frac{\bar{\rho}}{M^2}\frac{d\log{\sigma^{-1}}}{d\log{M}},\label{mf}
\end{equation}
where $f(\sigma)=2\sigma^2\mathcal{F}(\sigma^2)$ is the so called
``multiplicity function'' and $\mathcal{F}(S)=dF/dS$ is the
derivative of the fraction of mass in
halos with mass $>M[S]$. The goal of the excursion set theory is to calculate
$f(\sigma)$ from $\mathcal{F}(S)$ obtained as the first up-crossing 
distribution of the random walks $\delta(\textbf{x},R)$ 
as function of the smoothing radius $R[S]$ (or equivalently $S$, which plays the role of
pseudo-time). Trajectories starts at $S=0$
(i.e. large radii limit $R\rightarrow\infty$) with
$\delta_0=0$. The fraction of mass in halos is then inferred from counting
the rate at which trajectories hit for the first time an absorbing
barrier. The mass function at high redshifts can be inferred by simply rescaling
all linear quantities by the growth factor predicted by the cosmological model under consideration.

The nature of the random walks depends on the statistics
of the initial density field as well as the form of the filter function. 
In the case of a sharp-k filter, $W(k,R)=\theta(1-kR)$, the random
walks are Markovian with $\langle\delta(S)\delta(S')\rangle_c=\textrm{min}(S,S')$.\footnote{We use the underscore $c$ to indicated 
connected correlators.} In such a case the first-crossing
distribution is obtained by
solving a simple Fokker-Planck diffusion problem. 
However, the sharp-k filtering does not correspond to a
physical halo mass, since the filtered volume $V(R)$ is undefined 
(see e.g. discussion in \cite{MaggioreRiotto2010a}). The only
unambigously defined mass is associated with a top-hat
filter in real space, $W(x,R)=\theta(R-x)$ (sharp-x filter) 
whose Fourier transform reads as
\begin{equation} 
\tilde{W}(k,R)=\frac{3}{(kR)^3}[\sin(kR)-(kR)\cos(kR)].\label{sharpx}
\end{equation}
In such a case the evolution of the random walks acquires pseudo-time
correlations for which to our knowledge an analytical solution of the first-crossing distribution does not exist. 
Maggiore \& Riotto \cite{MaggioreRiotto2010a} have shown that such
correlations can be treated perturbatively around the Markovian
solution. In particular, for a sharp-x filter 
the two point connected correlator can
be written as $\langle\delta(S)\delta(S')\rangle_c=\textrm{min}(S,S')+\Delta(S,S')$,
where $\Delta(S,S')\approx \kappa \,S/S'(S'-S)<1$ (with $S<S'$)
and $\kappa$ is a small parameter of nearly constant amplitude, whose value 
depends on the underlying cosmology. For a vanilla $\Lambda$CDM model
$\kappa\approx 0.475$. Hence, the non-Markovian first-crossing
distribution can be inferred as a perturbative expansion in $\kappa$.

\subsection{Stochastic Barrier and Ellipsoidal Collapse}\label{stocmod}

\subsubsection{State of art of the stochastic barrier model}

The absorbing barrier is the other crucial ingredient of the excursion set 
formalism which encodes information on the non-linear collapse of halos.
In the spherical collapse model, the barrier is a constant threshold,
$B=\delta_c$. However, this model is a too simple description of the collapse
of dark matter halos in the universe. The gravitational collapse can be highly
non-spherical. As already shown in the seminal work by Doroshkevich \cite{Doroshkevich1970} perturbations 
in a Gaussian density field are triaxial and subject to external tidal shear. The implications of this are twofold. 
First, the collapse of a homogeneous ellipsoid \cite{EisensteinLoeb1995} provides a better description of the system. Secondly, the stochastic nature
of the underlying density field implies that the parameters characterizing the ellipsoid (ellipticity and prolatness) are random variables
with a well defined probability distribution that can be derived from that of the density field. Consequently,
the condition of collapse is no longer specified by a unique constant value, but it is a random variable itself \cite{Auditetal1997,Sheth2001}.
In the excursion set framework this implies treating the barrier as a stochastic variable which performs a random walk
whose characteristics are given by the moments of the probability distribution resulting from the ellipsoidal collapse model considered.
Unfortunatly the full distribution of the barrier remains unset. Most of the work dedicated to the ellipsoidal collapse threshold have limited their
analysis to the average behavior of the collapse threshold as function of the halo mass. As an example Sheth, Mo \& Tormen found that the 
average ellipsoidal collapse barrier evolves as $\langle B(S)\rangle=\delta_c (1+\beta S^\gamma)$. This behavior is consistent with the physical
expectation that collpase of small mass halos differ from spherical symmetry. In fact, at large $S$ (small mass) the shear field is higher
than at small $S$ (large mass) thus implying that small mass halos need to reach a higher density of collapse than 
that predicted by the spherical model. However, knowledge of the scatter around this average encodes additional information on the 
non-linear collapse process. To date the variance of the stochastic barrier has been studied only qualitatively \cite{Sheth2001}. 
In \cite{Desjacques2008} it has been shown that the dynamics of the collapse depends not only on the local properties of the shear field but also on interaction between ellipsoidal perturbations and its large scales environment. One of the key parameter to quantify this effect is the large scale environment density. Low mass haloes are more affected by this large scale environment and one of the consequence is to increase the scatter. This study was done for a Gaussian field but it already suggests that the imprint of the primordial density field is underline at hight mass halo while the dynamics and the cosmology dependences induce by the environment dynamics is underline in low mass haloes.\\

However we must not draw any conclusion on the value of these parameters without a proper account of the second order cloud-in-cloud problem. In the excursion set, we assume that we can consider random walks around the same coordinate holding on the ergodic principle. This hypothesis is approximative and it is possible to measure the induced discrepancy on the scatter by doing the analysis of \cite{Sheth2001}: comparing the scatter for center of mass walk and random particle of the simulation. Nevertheless this will require a dedicated study that we would like to perform in the future including non-Gaussian statistic of the primordial field. In the absence of prescription we focus a simpler model with captures the main feature of the fuzzy barrier.\\

\subsubsection{Modelisation of the Diffusing Drifting Barrier}

As we already introduce in \cite{PSCIA2011a,PSCIA2011b} we use a stochastic barrier with Gaussian diffusion and linear drifting
average, which we refer as the ``Diffusive Drifting Barrier'' model (DDB).
In this model the barrier performs a Markovian random walk
starting at $B_0=\delta_c$ with $\langle B(S)\rangle_c=\delta_c+\beta\,S$ and 
$\langle B(S)B(S')\rangle_c=D_B\,\textrm{min}(S,S')$.
Here $\beta$ parametrizes the rate at which the
collapse threshold on average deviates from the spherical collapse
prediction and $D_B$ measures the amplitude of a constant
scatter around the average collapse threshold at a given mass scale. The linear behaviour of the average of the threshold approximates the mean value of the Sheth Mo and Tormen ellispoidal barrier in the mass of interest.


Therefore, differently from the linear
density field $\delta$, the non-Markovian part of
the barrier 2-point connected correlator identically vanishes,
$\Delta_B(S,S')=0$. One may wonder if a better modeling of the stochastic barrier 
should account for the correlations induced by the filtering process. However, as already
pointed out in \cite{PSCIA2011b}, there is no reason as to why the
barrier should have the same filtering of the linear density field,
since the two procedures have very different physical meanings.
The latter is associated to the halo mass definition, while the former
specifies the correlation between the condition of collapse at
different scales. To date there is no study that has looked at such
correlation in the context of the ellipsoidal collapse model. In the lack
of information, the simplest approximation is to assume that the 
collapse condition at scale $S$ is independent of that at $S'$, i.e.
$\Delta_B(S,S')=0$. This suggests that in order to improve the
modeling of the stochastic barrier one should completely solve the
ellipsoidal collapse model through a detailed numerical analysis from
which to infer moments and correlators of the fuzzy barrier. This will also allow
to predict the redshift and cosmology dependence of $\beta$ and $D_B$.

The stochastic system can be simplified into one-dimensional random walks by
introducing the variable $Y=B-\delta$, 
with initial condition $Y_0=\delta_c$ and absorbing boundary at
$Y_c=0$. Non-Gaussian initial conditions can be easily included through
the higher-order connected correlation functions of the smoothed
linear density field. For instance, let us consider non-Gaussianity generated by the
three-point correlation function (skewness). In such a case the random walks are completely
specified by the non-vanishing connected correlators
\begin{eqnarray}
\langle Y(S)\rangle_c&=&\delta_c+\beta\,S,\label{avy}\\
\langle
Y(S)Y(S')\rangle_c&=&(1+D_B)\textrm{min}(S,S')+\Delta(S,S'),\label{vary}\\
\langle
Y(S)Y(S')Y(S'')\rangle_c&=&-\langle
\delta(S)\delta(S')\delta(S'')\rangle_c,\label{skewy}
\end{eqnarray}
where $\langle
\delta(S)\delta(S')\delta(S'')\rangle_c$ is the three point correlation function of the smoothed
linear density field. The goal of the excursion set is to compute
$\Pi(Y_0,Y,S)$, the probability distribution of the random
walks which start at $Y_0$ and reach $Y$ at time $S$ obeying eq.~(\ref{avy}),~(\ref{vary}) and~(\ref{skewy}), 
with $\Pi(Y_0,Y_c,S)=0$. Then,
the first-crossing distribution can be derived as
\begin{equation}
\mathcal{F}(S)=-\frac{\partial}{\partial{S}}\int^{\infty}_{Y_c} dY\,\Pi(Y_0,Y,S).\label{firstcross}
\end{equation}

\subsection{Path-Integral Formulation}
The computation of $\Pi(Y_0,Y,S)$ can be performed using
the path-integral method introduce in \cite{MaggioreRiotto2010a}. Let us consider a discretized interval  
$[0,S]$ in steps $\Delta{S}=\epsilon$, such that $S_k=k\epsilon$ with
$k=1,...,n$. The probability distribution of the discrete random walks
that never cross the barrier reads as 
\begin{equation}
\Pi_\epsilon(Y_0,Y_n,S_n)=\int^{\infty}_{Y_c} dY_1\,...\int^{\infty}_{Y_c} dY_{n-1} W(Y_0,..,Y_n,S_n),\label{pepsilon}
\end{equation}
where
\begin{equation}
W(Y_0,..,Y_n,S_n)\equiv\langle\delta_D(Y(S_1)-Y_1)...\delta_D(Y(S_n)-Y_n)\rangle,\label{wdensdirac}
\end{equation}
is the probability density distribution. Using the Fourier transform
of the Dirac-function we can rewrite eq.~(\ref{wdensdirac}) as
\begin{equation}
W(Y_0,..,Y_n,S_n)=\int \mathcal{D}\lambda\, 
  e^{i\sum_{i=1}^n\lambda_i Y_i}\langle e^{-i\sum_{i=1}^n\lambda_i Y(S_i)}\rangle,\label{wdens}
\end{equation}
with $\int\mathcal{D}\lambda=\int_{-\infty}^{\infty}\frac{d\lambda_1}{2\pi}...\frac{d\lambda_{n}}{2\pi}$.
As clearly pointed out in \cite{MaggioreRiotto2010c}, the exponential
term inside the average is nothing else than the exponential
of the partition function, $e^Z=\langle e^{-i\sum_{i=1}^n\lambda_i Y(S_i)}\rangle$.
For generic non-Gaussian random walks this reads as
\begin{equation}
Z=\sum_{p=1}^\infty\frac{(-i)^p}{p!}\sum_{i_1=1}^n...\sum_{i_p=1}^n\lambda_{i_1}...\lambda_{i_p}\langle
Y_{i_1}...Y_{i_p}\rangle_c,\label{zfunc}
\end{equation}
where $\langle Y_{i_1}...Y_{i_p}\rangle_c$ is the $p$-point connected
correlator. Using eq.~(\ref{avy}), (\ref{vary}) and (\ref{skewy}) and 
substituting in eq.~(\ref{wdensdirac}) we have
\begin{eqnarray} 
\Pi_\epsilon(Y_0,Y_n,S_n)&=&\int^{\infty}_{Y_c} dY_1\,...\int^{\infty}_{Y_c}
dY_{n-1}\int\mathcal{D}\lambda\,e^{-i\Sigma_k\lambda_k
  [\bar{B}_k-Y_k]-\frac{1}{2}\Sigma_{ij}\lambda_i\lambda_j A_{ij}}\times\nonumber\\ 
&\times&e^{-\frac{1}{2}\Sigma_{ij}\lambda_i\lambda_j\Delta_{ij}}e^{\frac{(-i)^3}{6}\Sigma_{ijk}\lambda_i\lambda_j\lambda_k\langle Y_i Y_j Y_k\rangle_c},\label{piepsexp}
\end{eqnarray}
with $A_{ij}=\epsilon (1+D_B)\textrm{min}(i,j)$. Since $\Delta_{ij}\equiv\Delta(S_i,S_j)<1$ and assuming
small departure from Gaussianity, we can expand the exponentials in eq.~(\ref{piepsexp})
to first order in $\Delta_{ij}$ and $\langle Y_i Y_j
Y_k\rangle_c$ respectively. Then using the relation $\lambda_i e^{-i\lambda_i [\bar{B}_i-Y_i]}=-i\partial/\partial{Y_i}
e^{-i\Sigma_{k}\lambda_k[\bar{B}_k-Y_k]}$ we can write 
\begin{equation}
\Pi_\epsilon(Y_0,Y_n,S_n)\approx \Pi^\epsilon_0(Y_0,Y_n,S_n)+\Pi^\epsilon_1(Y_0,Y_n,S_n)+\Pi^\epsilon_{NG}(Y_0,Y_n,S_n),\label{piexp}
\end{equation}
where
\begin{equation}
\Pi^\epsilon_0(Y_0,Y_n,S_n)=\int^{\infty}_{Y_c} dY_1\,...\int^{\infty}_{Y_c}
dY_{n-1} W_0(Y_0,...,Y_n,S_n),
\end{equation}
is the Markovian probability distribution with density
\begin{equation}
W_0(Y_0,...,Y_n,S_n)=\int\mathcal{D}\lambda\,e^{-i\Sigma_k\lambda_k[\bar{B}_k-Y_k]-\frac{1}{2}\Sigma_{ij}\lambda_i\lambda_j A_{ij}},
\end{equation}
which obeys the Chapman-Kolmogorov equation (see appendix A in ref.~\cite{PSCIA2011b}),
\begin{equation}
\Pi^\epsilon_1(Y_0,Y_n,S_n)=\frac{1}{2}\sum_{ij}\int^{\infty}_{Y_c} dY_1...\int^{\infty}_{Y_c}
dY_{n-1}\,\Delta_{ij}\partial_i\partial_j W_0(Y_0,..,Y_n,S_n),\label{nmcepsilon}
\end{equation}
is the first order correction in $\kappa$ and
\begin{equation}
\Pi^\epsilon_{NG}(Y_0,Y_n,S_n)=-\frac{1}{6}\sum_{ijk}\langle\delta_i\delta_j\delta_k\rangle_c\int^{\infty}_{Y_c}
dY_1...\int^{\infty}_{Y_c} dY_{n-1}\,\partial_i\partial_j\partial_k W_0(Y_0,..,Y_n,S_n)\label{ping}
\end{equation}
is the non-Gaussian term \cite{MaggioreRiotto2010c}. Notice that due to the separate expansions in
$\Delta_{ij}$ and $\langle Y_i Y_j Y_k\rangle_c$ we have mixing terms.
However, as we will discuss later, these are negligible to first order
in $\kappa$ and $f_{NL}$. By evaluating these integrals in the continuous limit
($\epsilon\rightarrow 0$) and computing the corresponding
first-crossing distribution we obtain a mass function that 
consists of three independent terms
\begin{equation}\label{ftot}
f(\sigma)=f_0(\sigma)+f_1(\sigma)+f_{NG}(\sigma),
\end{equation}
where $f_0(\sigma)$ is the Markovian part of the Gaussian mass
function, $f_1(\sigma)$ is the non-Markovian correction to the
Markovian solution to first order in $\kappa$ due to the filtering 
process of the linear density field and $f_{NG}(\sigma)$ is the
non-Gaussian contribution. Before presenting the computation of 
$f_{NG}(\sigma)$, it is worth reviewing the main
features of the Gaussian mass function derived in \cite{PSCIA2011a,PSCIA2011b}.

\subsection{Gaussian Halo Mass Function}\label{gaussmass}
The Markovian mass function for a diffusive drifting barrier reads as
\begin{equation}
f_0(\sigma)=\frac{\delta_c}{\sigma}\sqrt{\frac{2a}{\pi}}\,e^{-\frac{a}{2\sigma^2}(\delta_c+\beta\sigma^2)^2}.\label{fsigma0}
\end{equation}
where $a=1/(1+D_B)$. The non-Markovian correction due to the filter function is
given as an expasion to second order in $\beta$
\begin{equation}
f_1(\sigma)=f_{1,\beta=0}^{m-m}(\sigma)+f_{1,\beta^{(1)}}^{m-m}(\sigma)+f_{1,\beta^{(2)}}^{m-m}(\sigma)
\end{equation}
with
\begin{eqnarray}
f_{1,\beta=0}^{m-m}(\sigma)&=&-a\,\kappa \sqrt{\dfrac{2}{\pi}} \dfrac{Y_0\sqrt{a}}{\sigma}\left( e^{-\frac{a Y_0^2}{2\sigma^2}}-\frac{1}{2} \Gamma\left[0,\frac{aY_0^2}{2\sigma^2}\right]\right) ,\label{f1b0}\\
f_{1,\beta^{(1)}}^{m-m}(\sigma)&=&- a\,Y_0\,\beta\left(a\,\kappa\, \text{Erfc}\left[ Y_0\sqrt{\dfrac{a}{2\sigma^2}}\right]+ f_{1,\beta=0}^{m-m}(\sigma)\right),\label{f1b1}\\
f_{1,\beta^{(2)}}^{m-m}(\sigma)&=&a\,\kappa\frac{a\,\beta^2\sigma^2}{2}\biggl\{ 2\dfrac{a\,Y_0^2}{\sigma^2}\textrm{Erfc}\left[ Y_0\sqrt{\dfrac{a}{2\sigma^2}}\right]+\nonumber\\
&+& \frac{1}{\sqrt{\pi}} e^{-\dfrac{a\,Y_0^2}{2\sigma^2}}\left( \sqrt{\dfrac{a}{2\sigma^2}}Y_0-4\left( \dfrac{a}{2\sigma^2}\right)^{3/2} Y_0^3\right)+\nonumber\\
&+&\frac{1}{\sqrt{\pi}}\Gamma\left[0,\frac{a\,Y_0^2}{2\sigma^2}\right]\left(3\left(\dfrac{a}{2\sigma^2}\right)^{3/2} Y_0^3-\sqrt{\dfrac{a}{2\sigma^2}}Y_0\right)\biggr\},
\end{eqnarray}
where $Y_0=\delta_c$ and $\Gamma(0,z)$ is the incomplete
Gamma function. This expansion is justified by the fact that from the ellipsoidal collapse model we expect $\beta<1$ and
in \cite{PSCIA2011a,PSCIA2011b} we have shown that terms
$\mathcal{O}(>\beta^3)$ are negligible\footnote{In $f_{1,\beta^{(2)}}^{m-m}(\sigma)$ we had previously miss a contribution. However this term is negligeable for mass less than $10^{16}M_{sun}$ in \cite{PSCIA2011a,PSCIA2011b} } ,

\begin{figure}
\centering
\includegraphics[width=11cm]{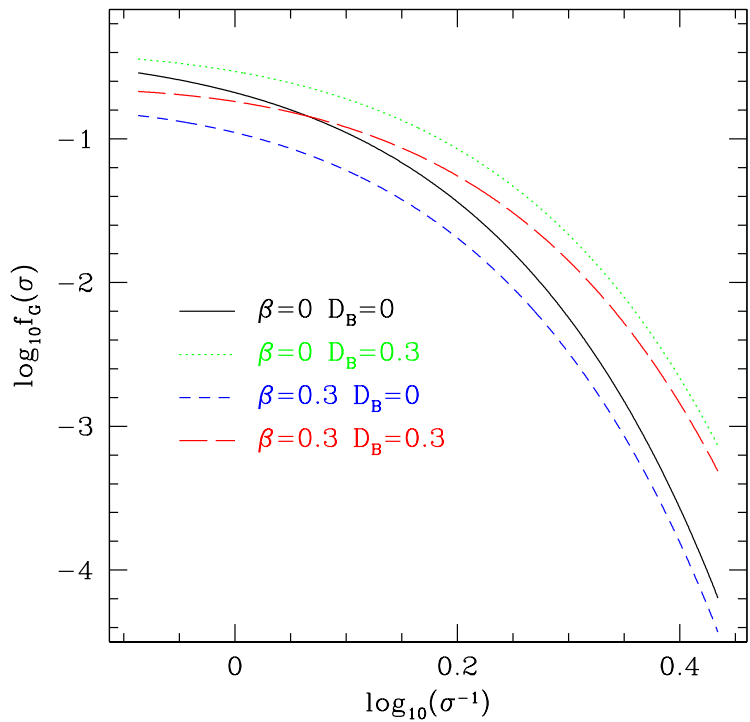} 
\caption{Gaussian excursion set halo mass function with sharp-x filter
  for the spherical collapse barrier (black solid line: $\beta=0$,
  $D_B=0$), a spherical diffusive
  barrier (green dot line: $\beta=0$, $D_B=0.3$), rigid moving
  barrier (blue short dash line: $\beta=0.3$, $D_B=0$) and diffusive drifting
  barrier (red long dash line: $\beta=0.3$, $D_B=0.3$).}\label{fig1}
\end{figure}

The collapse model affects the Gaussian mass function, $f_G(\sigma)=f_0(\sigma)+f_1(\sigma)$,
in a non-trivial manner (for an exhaustive discussion we refer the
reader to section III in ref.~\cite{PSCIA2011b}).
In figure~\ref{fig1} we plot $f_G(\sigma)$ 
for different values of $\beta$ and $D_B$ over the interval
corresponding to the mass range probed by current NG N-body
simulations ($10^{13}<M[\textrm{h}^{-1} M_\odot]<10^{15}$). The spherical 
collapse scenario studied in \cite{MaggioreRiotto2010a} corresponds to $\beta=0$ and
$D_B=0$. We can see that the effect of the average drift of the
barrier (see curve with $\beta=0.3$ and $D_B=0$) is to suppress 
the mass function at all masses with respect to the spherical collapse
case. Since the average threshold of collapse 
drifts toward larger values as function of $\sigma$, it becomes less likely 
for random walks to first-up crossing the barrier, thus decreasing
the amplitude of $f_G(\sigma)$ more effectively in the low-mass end
than at larger masses.
On the other hand the effect of the stochastic diffusion (see curve
with $\beta=0$ and $D_B=0.3$) is to facilitate the barrier crossing, thus 
effectively lowering the threshold of collapse. This is
consistent with the presence of the factor $a=1/(1+D_B)$ 
in the exponential cut-off of the mass function, which 
leads to a larger amplitude of the mass function especially in the
high-mass end.
In the case of a diffusing drifting barrier (see curve with $\beta=0.3$ and
$D_B=0.3$) these two effects compete to
produce a mass function which is tilted with respect to the spherical
collapse prediction. 

The inclusion of the
non-Markovian correction $f_1(\sigma)$ due to the filtering procedure has the
effect of decreasing the amplitude of the mass function. This is because in the presence
of pseudo-time correlations the random walks are characterized by
smoother trajectories with less frequent jumps. This has the effect of
disfavoring the first-up crossing of the barrier, thus lowering the mass function. 
This is consistent with the overall minus sign in eq.~(\ref{f1b0}) and
eq.~(\ref{f1b1}) which are the leading and next-to-leading non-Markovian
corrections in the mass range of interest. Here, it is also worth
noticing that these expressions carry higher powers of $a$ 
compared to the Markovian term $f_0(\sigma)$. 
This is because the effect of pseudo-time correlations on the mass
function is diluted in the presence of diffusion 
(since diffusion always favourates barrier crossing), thus leading to a faster
convergence of the expansion in $\kappa$. As shown in
\cite{PSCIA2011b}, the mass function with non-Markovian corrections to first order in
$\kappa$ is consistent within numerical errors with the exact numerical solution inferred 
from Monte Carlo generated random walks. 

The study presented in \cite{PSCIA2011a,PSCIA2011b} has shown that
implementing a realistic modeling of the non-spherical collapse of halos
in the excursion set formalism is key to inferring a
physical description of the halo mass function which is capable of reproducing
results from N-body to numerical accuracy. Here, we will show that
this is the case also for non-Gaussian initial conditions, since the
halo collapse model directly alters the non-Gaussian signal.
Previous works on the non-Gaussian halo mass function have simply
reabsorbed the effects of the ellipsoidal collapse of halos in the 
overall Gaussian normalization function, for instance using the
Sheth-Tormen formula \cite{ShethTormen1999}. But in doing so they have implicitly assumed that
the collapse is independent of the statistics of the primordial
density field and thus does not affect the non-Gaussian contribution
to the halo mass function. However, we have seen that the barrier model affects 
both the Markovian and non-Markovian part of the halo mass function. 
Since non-Gaussianity appears as a non-Markovian correction to the 
Gaussian term, it is reasonable to expect that the stochastic barrier
directly affects the dependence of the halo mass function on primordial
non-Gaussianity. We will show this in detail in the next sections.

\section{Non-Gaussian Halo Mass Function}\label{ngmass}
In order to compute eq.~(\ref{ping}) we need to first address the
scale dependence of the three-point correlator of the smoothed linear 
non-Gaussian density field. Following \cite{MaggioreRiotto2010c} we 
expand $\langle \delta_i \delta_j \delta_k\rangle_c$ in a triple Taylor series 
around $S_i=S_j=S_k=S_n$\footnote{In \cite{Damico2011} the authors compute
the path-integral through a saddle-point technique and identify scale
dependent small parameters which allow them to infer a mass function
which is valid over a larger range of masses.}: 
\begin{equation}\label{d3exp}
\langle \delta(S_i) \delta(S_j)
\delta(S_k)\rangle_c=\sum_{p,q,r=0}^\infty
\frac{(-1)^{p+q+r}}{p!q!r!}(S_n-S_i)^p (S_n-S_j)^q (S_n-S_k)^r G_{3}^{(p,q,r)}(S_n),
\end{equation}
where
\begin{equation}
 G_{3}^{(p,q,r)}(S_n)\equiv\frac{d^p}{dS_i^p}\frac{d^q}{dS_j^q}\frac{d^r}{dS_k^r}\langle \delta(S_i) \delta(S_j)
\delta(S_k)\rangle_c\bigg|_{S_i=S_j=S_k=S_n}.
\end{equation}
In the large mass limit ($S_n\rightarrow 0$), where we expect that the
primordial NG signature to be less altered by non-linear effects, the leading contribution
is given by the term $\langle\delta^3(S_n)\rangle$, corresponding to
$p+q+r=0$. The next-to-leading order is given by $p+q+r=1$
corresponding to three terms proportional to derivative of
$\langle\delta(S_i)\delta^2(S_n)\rangle$ and so on, thus
\begin{equation}
\Pi^\epsilon_{NG}(Y_0,Y_n,S_n)=\Pi^{\epsilon,L}_{NG}(Y_0,Y_n,S_n)+\Pi^{\epsilon,NL}_{NG}(Y_0,Y_n,S_n)+...
\end{equation} 
In \cite{DeSimone2011}, the authors have computed the excursion set 
non-Gaussian mass function for a generic rigidly moving barrier modeled as a Taylor
expansion in $S_n$ up to leading order. Here, we provide an exact
computation for the stochastic barrier with linearly drifting average
up to next-to-leading order. 

\subsection{Leading Term}
The zero order term $p=q=r=0$ reads as
\begin{equation}\label{pinglo}
\Pi^{\epsilon,L}_{NG}(Y_0,Y_n,S_n)=-\frac{1}{6}\langle\delta^3(S_n)\rangle\sum_{i,j,k=1}^{n}\int^{\infty}_{Y_c}
dY_1...\int^{\infty}_{Y_c} dY_{n-1}\,\partial_i\partial_j\partial_k W_0(Y_0,..,Y_n,S_n),
\end{equation}
We can decompose the sum as
\begin{equation}\label{sumdec}
\sum_{i,j,k=1}^{n}\partial_i\partial_j\partial_k=\partial_n^3+3\sum_{i=1}^{n-1}\partial_n^2\partial_i+3\sum_{i,j=1}^{n-1}\partial_n\partial_i\partial_j+
\sum_{i,j,k=1}^{n-1}\partial_i\partial_j\partial_k,
\end{equation}
the integrals associated to these terms can be computed using the fact that
\begin{equation}
\frac{\partial}{\partial
  Y_c}\Pi_0^\epsilon(Y_0,Y_n,S_n)=-\sum_{i=1}^{n-1}\int_{Y_c}^\infty
  dY_1...\int_{Y_c}^\infty dY_{n-1}\partial_i W_0(Y_0,..,Y_n,S_n),
\end{equation}
\begin{equation}
\frac{\partial^2}{\partial
  Y_c^2}\Pi_0^\epsilon(Y_0,Y_n,S_n)=\sum_{i,j=1}^{n-1}\int_{Y_c}^\infty
  dY_1...\int_{Y_c}^\infty dY_{n-1}\partial_i\partial_j W_0(Y_0,..,Y_n,S_n),
\end{equation}
and 
\begin{equation}\label{pi3lo}
\frac{\partial^3}{\partial Y_c^3}\Pi_0^\epsilon(Y_0,Y_n,S_n)=-\sum_{i,j,k=1}^{n-1}\int^{\infty}_{Y_c}
dY_1...\int^{\infty}_{Y_c} dY_{n-1}\,\partial_i\partial_j\partial_k W_0(Y_0,..,Y_n,S_n).
\end{equation}
The first-crossing distribution is given by eq.~(\ref{firstcross})
and one can easily show that the integrals in $dY_n$ (in the
continous limit) of the first three terms in eq.~(\ref{sumdec})  
vanish. Thus, the only non-vanishing contribution to the first-crossing distribution
is given by the term in eq.~(\ref{pi3lo}) such
that:\footnote{Following the same 
derivation it is easy to obtain the non-Gaussian contribution to the
halo mass function from a non-vanishing 4-point correlation function by expanding the
path-integral eq.~(\ref{wdens}) to first order in
$\langle\delta_i\delta_j\delta_k\delta_k\rangle$. Then, to leading
order in $S_i=S_j=S_k=S_l=S_n$ the first crossing distribution is 
$-\frac{1}{4!}\frac{\partial}{\partial S}\left[\langle\delta^4(S)\rangle\frac{\partial^4}{\partial
    Y_c^4}\int_{Y_c}^\infty dY \,\Pi_0(Y_0,Y,S)\right]$.}
\begin{equation}\label{fracngl}
\mathcal{F}_{NG}^L(S)=\frac{1}{6}\,\frac{\partial}{\partial
  S}\left[\langle\delta^3(S)\rangle\frac{\partial^3 I_L}{\partial Y_c^3}\bigg|_{0}\right],
\end{equation}
where
\begin{equation}\label{IL}
I_L=\int^{\infty}_{Y_c}\Pi_0(Y_0,Y_c,Y,S)\,dY.
\end{equation}
This integral can be computed exactly using the Markovian solution \cite{PSCIA2011b}
\begin{equation}
\Pi_0(Y_0,Y_c,Y,S)=\sqrt{\frac{a}{2\pi S}}e^{a\beta\left(Y-Y_0-\beta
  S/2\right)}\left[e^{-\frac{a}{2S}(Y-Y_0)^2}-e^{-\frac{a}{2S}(2 Y_c-Y-Y_0)^2}\right],
\end{equation}
obtaining
\begin{equation}
I_L=\frac{1}{2}\left\{\textrm{Erfc}\left(\frac{Y_c-Y_0-\beta
  S}{\sqrt{2
    S(1+D_B)}}\right)-e^{2\beta\frac{Y_c-Y_0}{1+D_B}}\textrm{Erfc}\left(\frac{Y_0-Y_c-\beta S}{\sqrt{2 S(1+D_B)}}\right)\right\}.
\end{equation}
Substituing this expression in eq.~(\ref{fracngl}) and using the fact that
$f(\sigma)=2\sigma^2\mathcal{F}(\sigma)$
we finally obtain 
\begin{equation}\label{fngl}
\begin{split}
f_{NG}^L(\sigma)&=\sqrt{\dfrac{2}{\pi}}e^{-\dfrac{a(Y_0+\beta\sigma^2)^2}{2\sigma^2}}\frac{a^{3/2}\sigma}{6} \left\lbrace S_3(\sigma)\left[  \dfrac{a^2}{\sigma^4}Y_0^4-2\dfrac{a}{\sigma^2}Y_0^2-1+3\dfrac{a^2}{\sigma^2}\beta Y_0^3+\right.\right.\\
&\left.\left.+3 a\beta Y_0+a^2\beta^3\sigma^2 Y_0+3 a^2(\beta Y_0)^2+13 a \beta^2\sigma^2\right]+\right.\\
&\left.+\dfrac{dS_3(\sigma)}{d\text{ln}\sigma}\left[ \dfrac{a}{\sigma^2}Y_0^2-1+3 a\beta Y_0+4 a\beta^2\sigma^2\right] \right\rbrace +\\
&+e^{-2a\beta Y_0}\text{Erfc}\left[\sqrt{\dfrac{a}{2\sigma^2}}(Y_0-\beta\sigma^2)\right]\dfrac{2a^{3}\sigma}{3}\beta^3\sigma^3\left\lbrace 4 S_3(\sigma)+\dfrac{dS_3(\sigma)}{d\text{ln}\sigma}\right\rbrace 
\end{split}
\end{equation}
where we have introduced the reduced cumulant
$\mathcal{S}_3(\sigma)=\langle\delta^3(S)\rangle\,/\sigma^4$. 
For the spherical collapse barrier, i.e. $\beta=0$ and $a=1$ ($D_B=0$), the
above expression reduces to Loverde et al. \cite{Loverdeetal2008} formula
as already noticed in \cite{MaggioreRiotto2010c}. On the other hand, it is worth
noticing the presence of additional mixing terms proportional to
powers of $\beta$ which couple the effects of the non-spherical halo
collapse to primordial non-Gaussianity.

\subsection{Next-to-Leading Term}
The next term in the expansion eq.~(\ref{d3exp}) correspond to $p+q+r=1$ (i.e. $p=1$ and $q=r=0$
plus permutations)
\begin{equation}\label{pinl}
\Pi^{\epsilon,NL}_{NG}(Y_0,Y_n,S_n)=3\times\frac{1}{6}G_3^{(1,0,0)}(S_n)\sum_{i=1}^n(S_n-S_i)\sum_{j,k=1}^n\int^{\infty}_{Y_c}
dY_1...\int^{\infty}_{Y_c} dY_{n-1}\partial_i\partial_j\partial_k W_0,
\end{equation}
where\footnote{We have $G_3^{(1,0,0)}(S_n)=G_3^{(0,1,0)}(S_n)=G_3^{(0,0,1)}(S_n)$
  since the $3$-point correlation function 
  $\langle\delta_i \delta_j \delta_k\rangle$ is symmetric under
  permutation of the indeces, a property that results of the fact that
  its dependence on $S_i$, $S_j$ and $S_k$ is given by the product of
  three identical filter functions computed at $R_i$, $R_j$ and $R_k$ respectively.} 
\begin{equation}
G_3^{(1,0,0)}(S_n)\equiv\left[\frac{d}{d S_i}\langle\delta(S_i)^2\delta(S_n)\rangle\right]_{S_i=S_n}.
\end{equation}
Again, decomposing the sum and using $\sum_{j,k=1}^{n-1}\rightarrow\partial^2/\partial Y_c^2$, 
we obtain
\begin{equation}\label{intpinl}
\int_{Y_c}^\infty dY_n\,
\Pi^{\epsilon,NL}_{NG}(Y_0,Y_n,S_n)=\frac{1}{2}G_3^{(1,0,0)}(S_n)\sum_{i=1}^n(S_n-S_i)\frac{\partial^2}{\partial
Y_c^2}\biggl[\int^{\infty}_{Y_c}
dY_1...\int^{\infty}_{Y_c} dY_{n}\,\partial_i W_0\biggr].
\end{equation}
The multiple integral in the above expression can be computed by
part and using the fact that $W_0$ obeys the Chapman-Kolmogorv
equation we can rewrite the integrand as
\begin{equation}
W_0(Y_0,..,Y_{i-1},Y_i=Y_c,Y_{i+1},...,Y_n,S_n)=W_0(Y_0,..,Y_c,S_i)W_0(Y_c,...,Y_n,S_n-S_i).
\end{equation}
After taking the continous limit, the
first-crossing distribution at next-to-leading order reads as
\begin{equation}
\mathcal{F}_{NG}^{NL}(S)=-\frac{1}{2}\frac{\partial}{\partial
  S}\left[G_3^{(1,0,0)}(S)\frac{\partial^2I_{NL}}{\partial Y_c^2}\bigg|_{0}\right],
\end{equation}
where
\begin{equation}\label{inl}
I_{NL}=-\int_0^S
dS' (S-S')\Pi^{f}_{0}(Y_0,Y_c,S')\int_{Y_c}^\infty dY\,\Pi^{f}_{0}(Y_c,Y,S-S'),
\end{equation}
with $\Pi^{f}_{0}(Y_0,Y_c,S)$ and $\Pi^{f}_{0}(Y_c,Y,S-S')$ are the
finite pseudo-time corrections at the barrier location (see \cite{PSCIA2011b}):
\begin{eqnarray}
\Pi^{f}_{0}(Y_0,Y_c,S)&=&\frac{a}{\sqrt{\pi}S^{3/2}}(Y_c-Y_0)e^{-\frac{a}{2 S}(Y_0-Y_c+\beta
    S)^2}\\
\Pi^{f}_{0}(Y_c,Y,S)&=&\frac{a}{\sqrt{\pi}S^{3/2}}(Y-Y_c)e^{-\frac{a}{2 S}(Y-Y_c-\beta
    S)^2},
\end{eqnarray}
eq.~(\ref{inl}) can be computed analytically using the fact that 
\begin{equation}
\int_0^S dS_i \frac{e^{-\frac{a^2}{2S_i}}e^{-\frac{b^2}{2(S-S_i)}}}{S_i^{1/2}(S-S_i)^{3/2}}=\frac{1}{b}\sqrt{\frac{2\pi}{S}}e^{-\frac{1}{2S}(a+b)^2},
\end{equation}
we obtain
\begin{equation}
I_{NL}=a(\beta \sigma^2+Y_c-Y_0)e^{2a\beta(Y_c-Y_0)}\text{Erfc}\left[\sqrt{\dfrac{a}{2\sigma^2}}(Y_0-Y_c-\beta\sigma^2)\right]+\sqrt{\dfrac{2}{\pi}}\sqrt{a}\sigma e^{-a\dfrac{(Y_0-Y_c+\beta\sigma^2)^2}{2\sigma^2}}
\end{equation}

Introducing $\mathcal{U}_3(S)\equiv\frac{3}{S}\,G_3^{(1,0,0)}(S)$ we finally obtain the multiplicity function to next-leading order
\begin{equation}\label{fngnl}
\begin{split}
f_{NG}^{NL}(\sigma)&=\dfrac{-1}{\sqrt{2\pi}}e^{-a\dfrac{(Y_0+\beta\sigma^2)^2}{2\sigma^2}}\dfrac{a^{3/2}\sigma}{3}\\
&\Biggl\{ U_3(\sigma)\left[1+\dfrac{a}{\sigma^2}Y_0^2+15a\beta^2\sigma^2+4a\beta Y_0\right]+\dfrac{dU_3(\sigma)}{d\ln\sigma}\left[1+4a\beta^2\sigma^2\right]\Biggr\} -\\
&-e^{-2a\beta Y_0}\text{Erfc}\left[\sqrt{\dfrac{a}{2\sigma^2}}(Y_0-\beta\sigma^2)\right]\dfrac{a^{3/2}\sigma}{3}\\
&\Biggl\{U_3(\sigma)\left[4\sqrt{a}\beta\sigma+8 (a)^{3/2}(\beta\sigma)^3-4 a^{3/2}\beta^2\sigma Y_0\right]+\\
&+\dfrac{dU_3(\sigma)}{d\text{ln}\sigma}\left[2\sqrt{a}\beta\sigma+2 a^{3/2}(\beta\sigma)^3-2 a^{3/2}\beta^2\sigma Y_0\right]\Biggr\}
\end{split}
\end{equation}

As in the case of the leading order term we notice the presence of additional mixing terms in powers of
$\beta$. This clearly indicates that contrary to standard
derivations of the NG mass function based on the Press-Schecther
formalism, the effect of the non-spherical collapse of halos cannot
be simply reabsorbed in the Gaussian part of the mass function, but directly
alters the non-Gaussian dependence\footnote{Notice that the
    non-Gaussian terms for $\beta=0$ do not exactly coincide with the
    corresponding formulae in \cite{MaggioreRiotto2010c} neglecting
    the mixing terms in $a \kappa$ and $f_{NL}$. The mass function for
    a diffusive stochastic barrier cannot be inferred from the
    spherical collapse scale by a simple rescaling of the
    variables including the skewness of the initial density field.}.

\section{Local and Equilateral Non-Gaussian Mass Function}\label{locequi}
\subsection{Reduced bispectra and fitting functions} 
In deriving the non-Gaussian terms we have
made no assumptions on the specific type of non-Gaussianity.
In the case of non-Gaussianity with non-vanishing skewness
the three-point correlation function of the smoothed linear
density field at $z=0$ formally reads as:
\begin{eqnarray}\label{d3R1R2R3}
\langle\delta(R_1)\delta(R_2)\delta(R_3)\rangle_c&=&\int\frac{d^3
  k_1}{(2\pi)^3}\int\frac{d^3 k_2}{(2\pi)^3}\int\frac{d^3
  k_3}{(2\pi)^3}\tilde{W}(k_1,R_1)\tilde{W}(k_2,R_2)\tilde{W}(k_3,R_3)\times\nonumber\\
&\times&\mathcal{M}(k_1)\mathcal{M}(k_2)\mathcal{M}(k_3)\langle\zeta(\textbf{k}_1)\zeta(\textbf{k}_2)\zeta(\textbf{k}_3)\rangle_c,
\end{eqnarray}
where
\begin{equation}
\mathcal{M}(k)=\frac{2}{5 H_0^2\Omega_m}T(k)k^2,
\end{equation}
with $H_0$ being the Hubble constant, $\Omega_m$ the matter density relative
to the critical one, $T(k)$ the transfer function at $z=0$
and $\zeta(k)$ is the curvature perturbation with
\begin{equation}
\langle\zeta(\textbf{k}_1)\zeta(\textbf{k}_2)\zeta(\textbf{k}_3)\rangle_c=(2\pi)^3\delta_D(\textbf{k}_1+\textbf{k}_2+\textbf{k}_3)B_{\zeta}(k_1,k_2,k_3),
\end{equation}
where $B_{\zeta}(k_1,k_2,k_3)$ is the reduced bispectrum and encodes
the characteristic momentum dependence of the non-Gaussian signal. 
The dependence on the Dirac-$\delta$ function, $\delta_D(\textbf{k})$,
ensures the momentum conservation, implying that admitted momentum
space configurations form closed triangles.

A variety of models, including standard inflation, predicts
\begin{equation}
B_{\zeta}(k_1,k_2,k_3)=\frac{6}{5}f^{\textrm{loc}}_{NL}\left[P_\zeta(k_1)P_\zeta(k_2)+P_\zeta(k_1)P_\zeta(k_3)+P_\zeta(k_2)P_\zeta(k_3)\right],
\end{equation}
where $f^{\textrm{loc}}_{NL}$ is a constant amplitude and
$P_\zeta(k)=A_s k^{n_s-4}$ is the primordial power
spectrum. This type of non-Gaussianity is also known as ``local'' since
the curvature perturbation at any point in space can be written in
terms of a linear Gaussian field, $\zeta_G(\textbf{x})$:
\begin{equation}
\zeta(\textrm{x})=\zeta_G(\textbf{x})+\frac{3}{5}f_{NG}^\textrm{loc}\left[\zeta_G^2(\textbf{x})-\langle\zeta^2_G(\textbf{x})\rangle\right].
\end{equation}
In such a case the non-Gaussian signal peaks around
momentum space configurations for which one of the momenta is much
smaller than the others (e.g. $k_1\ll k_2\approx k_3$). WMAP-7yr
measurements of the CMB bispectrum constrain the amplitude
of local non-Gaussianity in the range $-10<f_{NG}^\textrm{loc}<74$ at
$2\sigma$ \cite{Komatsu2011}.

Alternatively, models in which primordial density fluctuations 
are generated by a field subject to higher-derivative terms
predict a reduced bispectrum of the form
\begin{eqnarray}
B_{\zeta}(k_1,k_2,k_3)&=&\frac{18}{5}f^{\textrm{equi}}_{NL}A_s^2\biggl[\frac{1}{2
    k_1^{4-n_s}k_2^{4-n_s}}+\frac{1}{3(k_1 k_2
    k_3)^{2(4-n_s)/3}}+\nonumber\\
&-&\frac{1}{(k_1 k_2^2
    k_3^3)^{(4-n_s)/3}}+\textrm{5 perm.}\biggr].
\end{eqnarray}
This is also known as ``equilateral'' non-Gaussianity, since the signal maximizes in
momentum space configurations for which $k_1\approx k_2 \approx k_3$. 
WMAP-7yr measurements constrain the amplitude of equilateral
non-Gaussianity in the range $-214<f_{NL}^{\textrm{equi}}<266$ at
$2\sigma$ \cite{Komatsu2011}.

\begin{figure}
\centering
\begin{tabular}{cc}
\includegraphics[width=7cm]{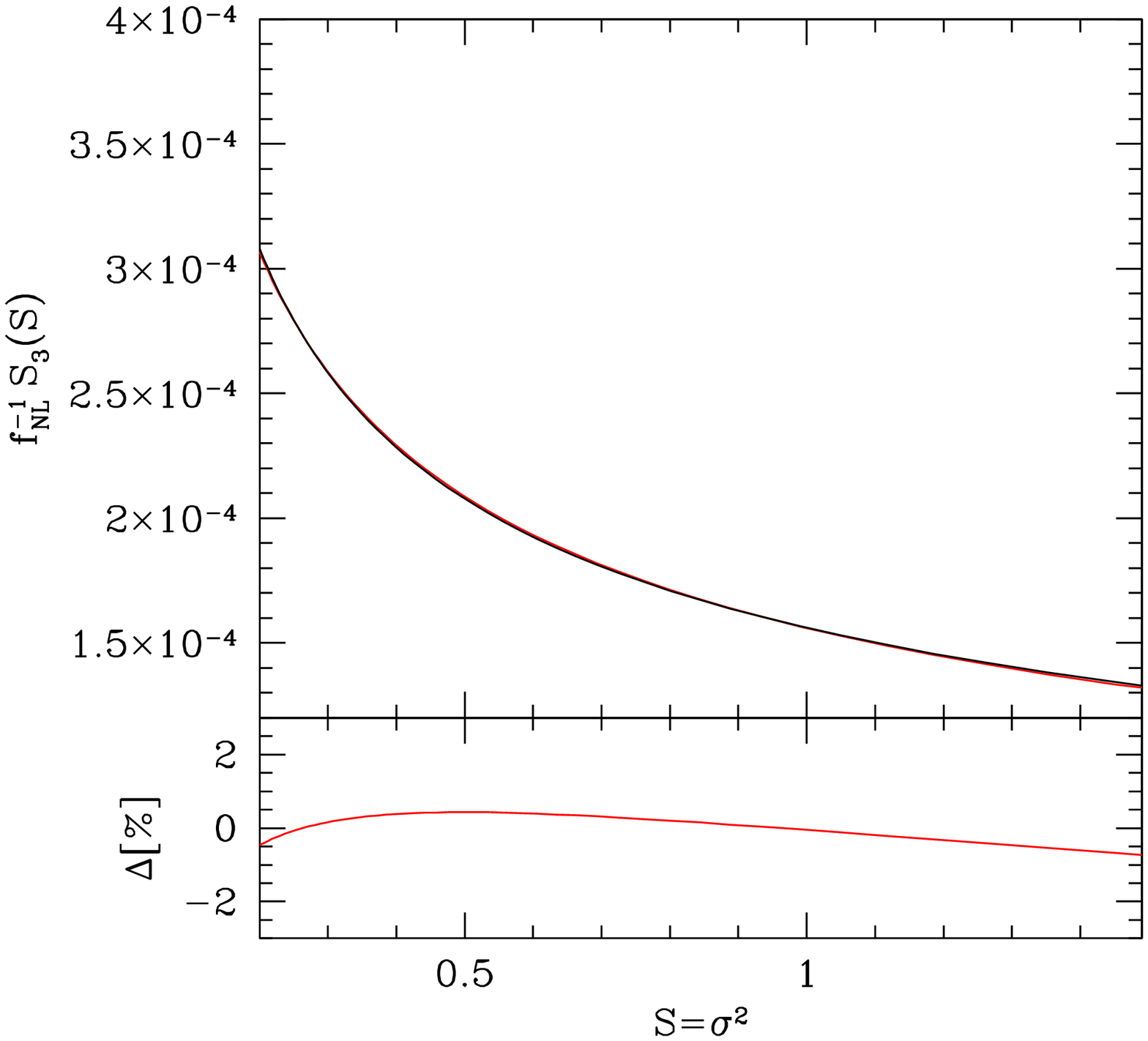} &
\includegraphics[width=7cm]{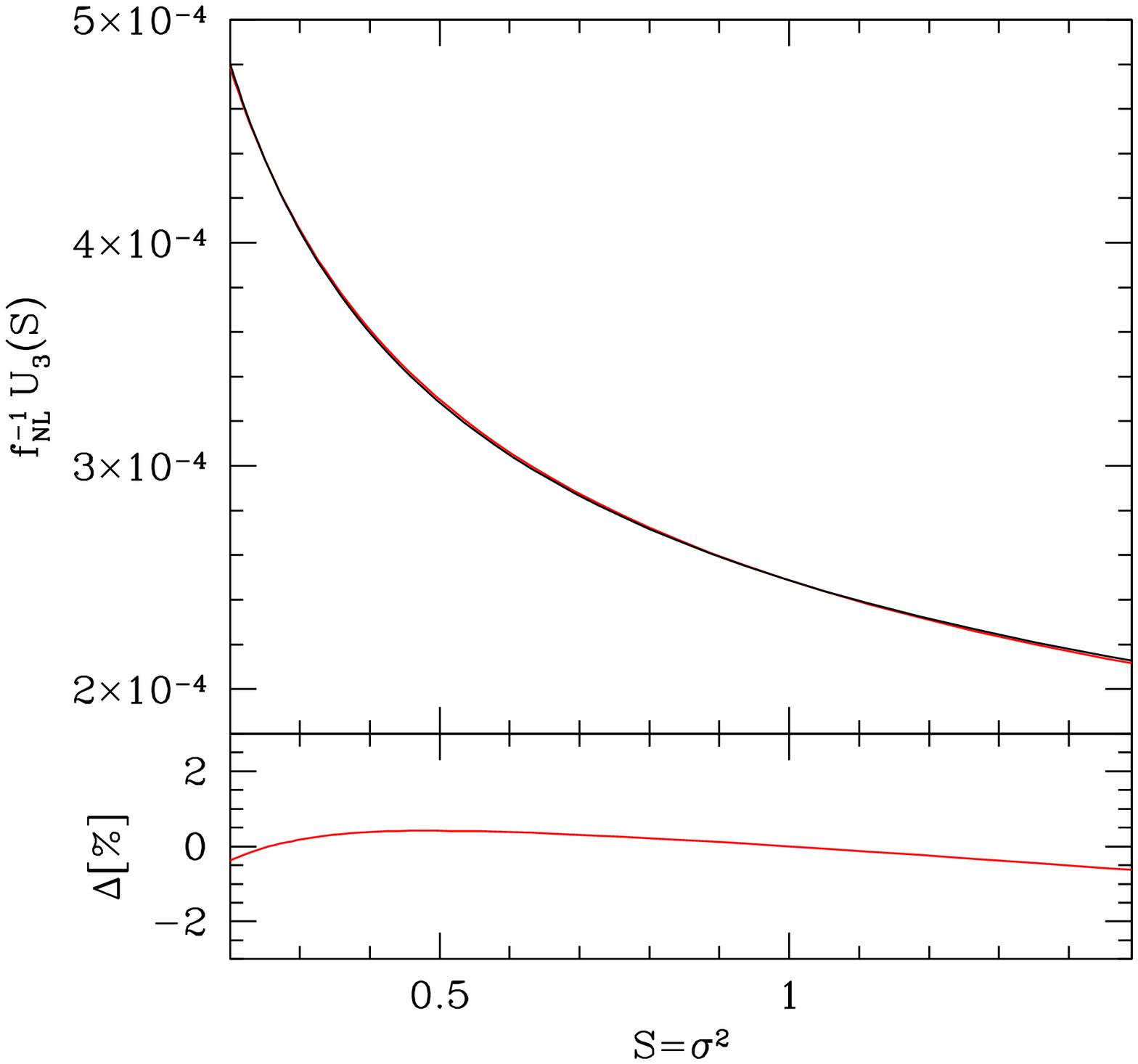}
\end{tabular}
\caption{$\mathcal{S}_3$ (a) and $\mathcal{U}_3$ (b) 
for local non-Gaussianity. Black
  solid curves are the numerically computed functions obtained from numerical
  integration of eq.~(\ref{d3R1R2R3}). Red solid curves correspond to
  fitting formulae eqs.~(\ref{s3locfit}),(\ref{u3fitloc}) with a few
  percent errors (bottom panels).}\label{fig2}
\end{figure}

\begin{figure}
\centering
\begin{tabular}{cc}
\includegraphics[width=7cm]{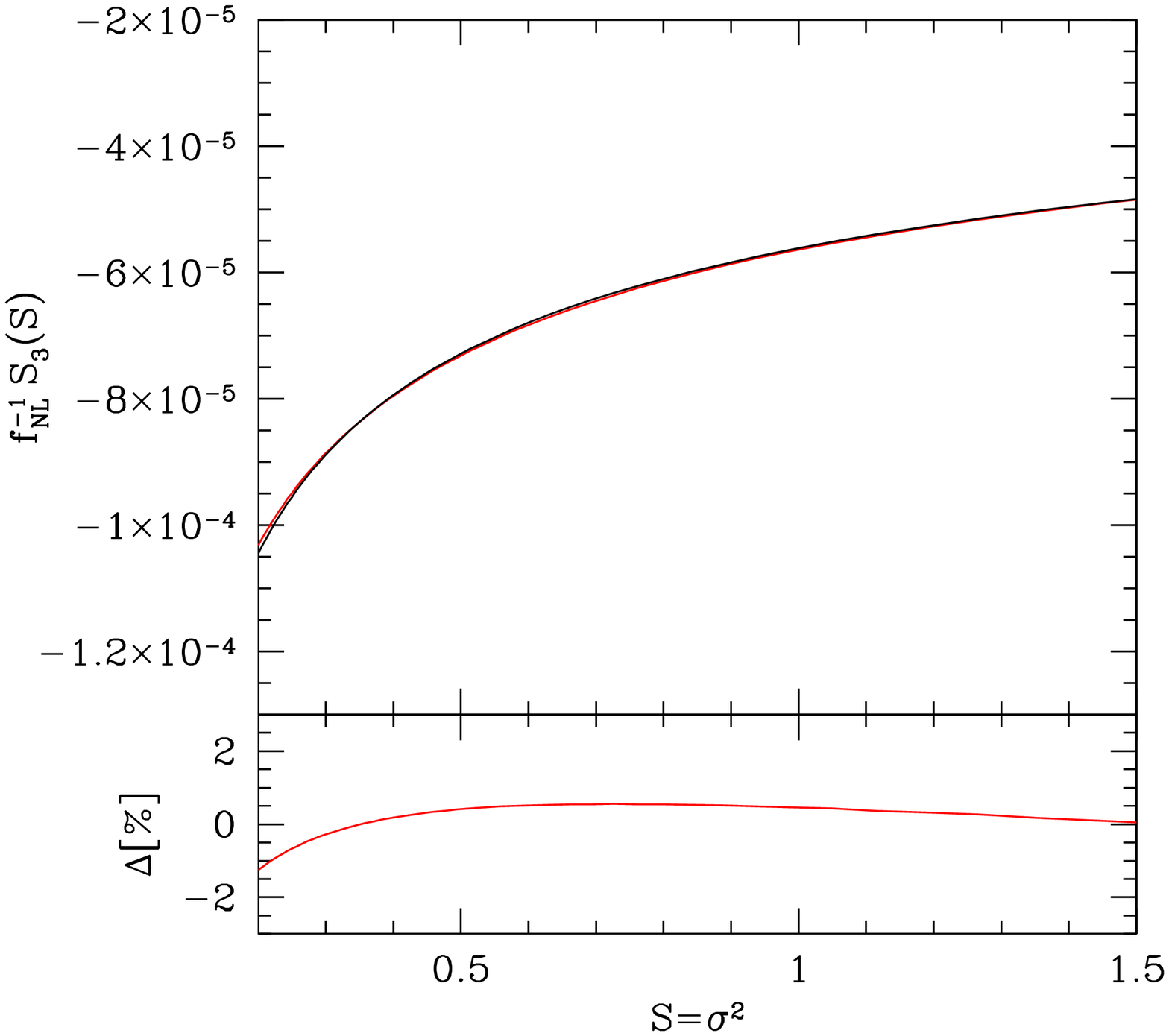} &
\includegraphics[width=7cm]{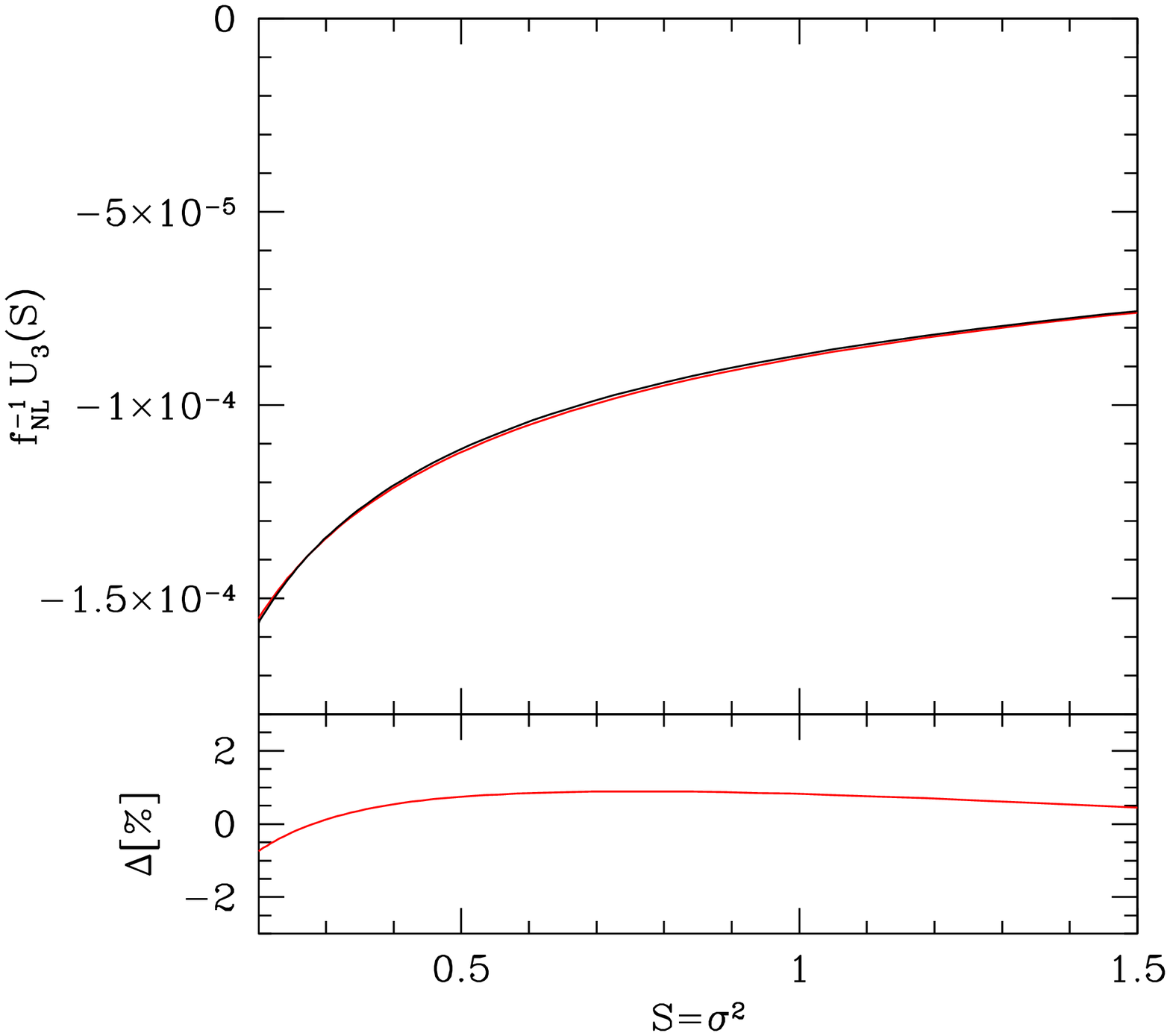}
\end{tabular}
\caption{As in figure~\ref{fig2} for equilateral
  non-Gaussianity. Fitting formulae are given by eqs.~(\ref{s3fitequi})-(\ref{u3fitequi})}\label{fig3}
\end{figure}

We integrate eq.~(\ref{d3R1R2R3}) numerically for local and
equilateral type of non-Gaussianities and infer
$\mathcal{S}_3$, $d\,\mathcal{S}_3/d\ln\sigma$, $\mathcal{U}_3$ 
and $d\,\mathcal{U}_3/d\ln\sigma$. The transfer function has been
computed for a vanilla $\Lambda$CDM model best fit to WMAP-7 years
data using the CAMB code \cite{CAMB}. The functions are plot 
in figure~\ref{fig2} and ~\ref{fig3} respectively, 
over the mass range probed by ~\cite{Pillepich10}, which corresponds 
to $-0.2\lesssim\ln(1/\sigma)\lesssim 0.8$.
These functions can be approximated to
better than a few percent accuracy by simple fitting formulae. In the
local case we find

\begin{equation}
\mathcal{S}^{\textrm{loc}}_3(S)=f_{NL}\dfrac{1.56}{S^{0.42}}10^{-4}
\label{s3locfit}
\end{equation}

\begin{equation}
\frac{d\mathcal{S}^{\textrm{loc}}_3}{dS}=-f_{NL}\dfrac{6.314985}{S^{1.4736}}10^{-5}
\label{ds3locfit}
\end{equation}

\begin{equation}
\mathcal{U}^{\textrm{loc}}_3(S)=f_{NL}\dfrac{2.487}{S^{0.407}}10^{-4}
\label{u3fitloc}
\end{equation}

\begin{equation}
\frac{d\,\mathcal{U}^{\textrm{loc}}_3}{dS}=-f_{NL}\dfrac{1.0122}{S^{1.4}}10^{-4}
\label{du3fitloc}
\end{equation}

while for equilateral NG we have
\begin{equation} 
\mathcal{S}^{\textrm{equi}}_3(S)=-f_{NL}\dfrac{5.4347}{S^{0.46}}10^{-5}
\label{s3fitequi}
\end{equation}

\begin{equation}
\frac{d\mathcal{S}^{\textrm{equi}}_3}{dS}=f_{NL}\dfrac{2.5}{S^{1.46}}10^{-5}
\label{ds3fitequi}
\end{equation}

\begin{equation}
\mathcal{U}^{\textrm{equi}}_3(S)=-f_{NL}\dfrac{8.78}{S^{0.354}}10^{-5}
\label{u3fitequi}
\end{equation}

\begin{equation}
\frac{d\,\mathcal{U}^{\textrm{equi}}_3}{dS}=f_{NL}\dfrac{3.108}{S^{1.354}}10^{-5}
\label{du3fitequi}
\end{equation}

\subsection{Validity of the perturbative expansions: limits on $f_{NL}$}
\begin{figure}
\centering
\begin{tabular}{cc}
\includegraphics[width=9cm]{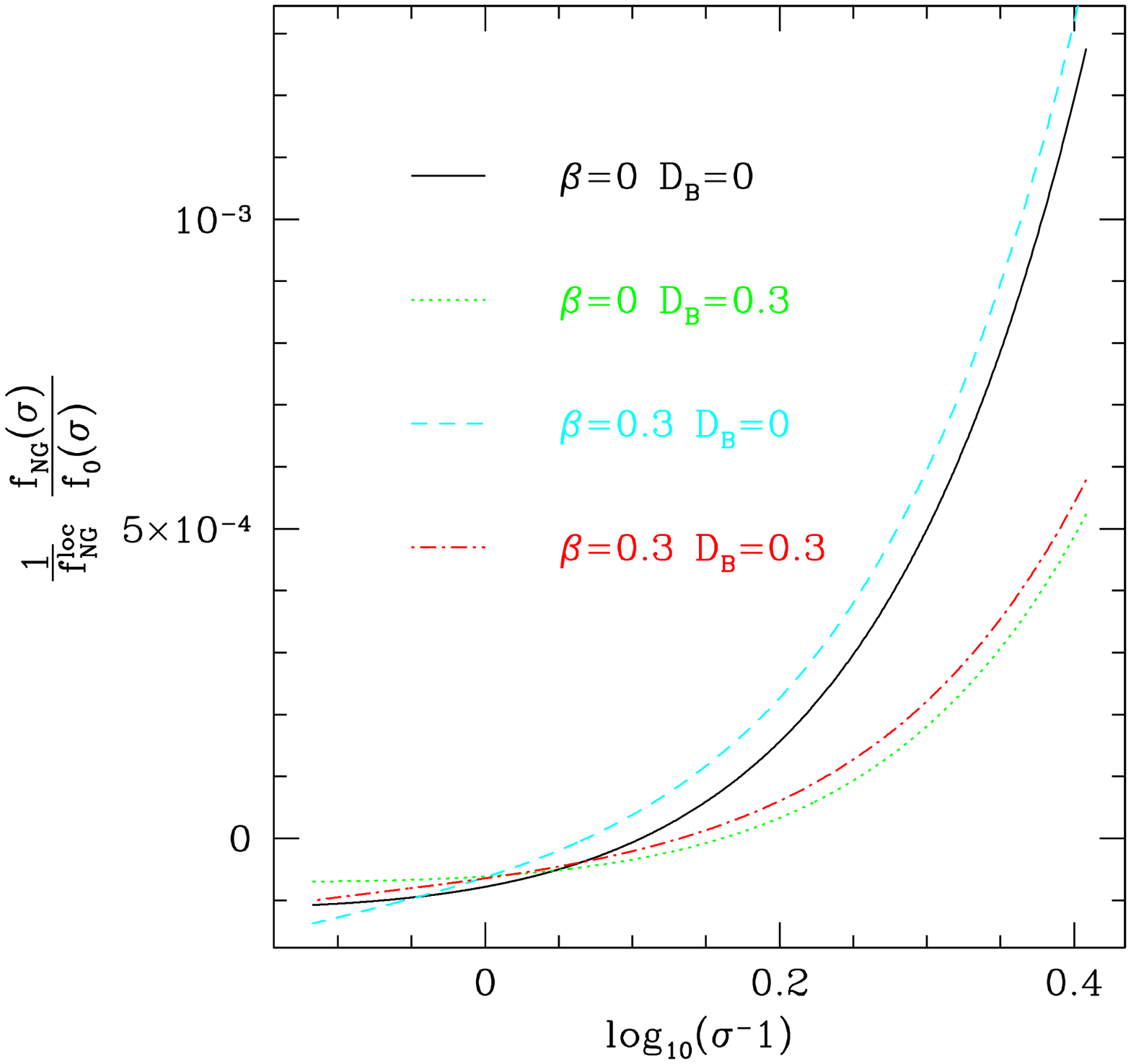} \\

\includegraphics[width=9cm]{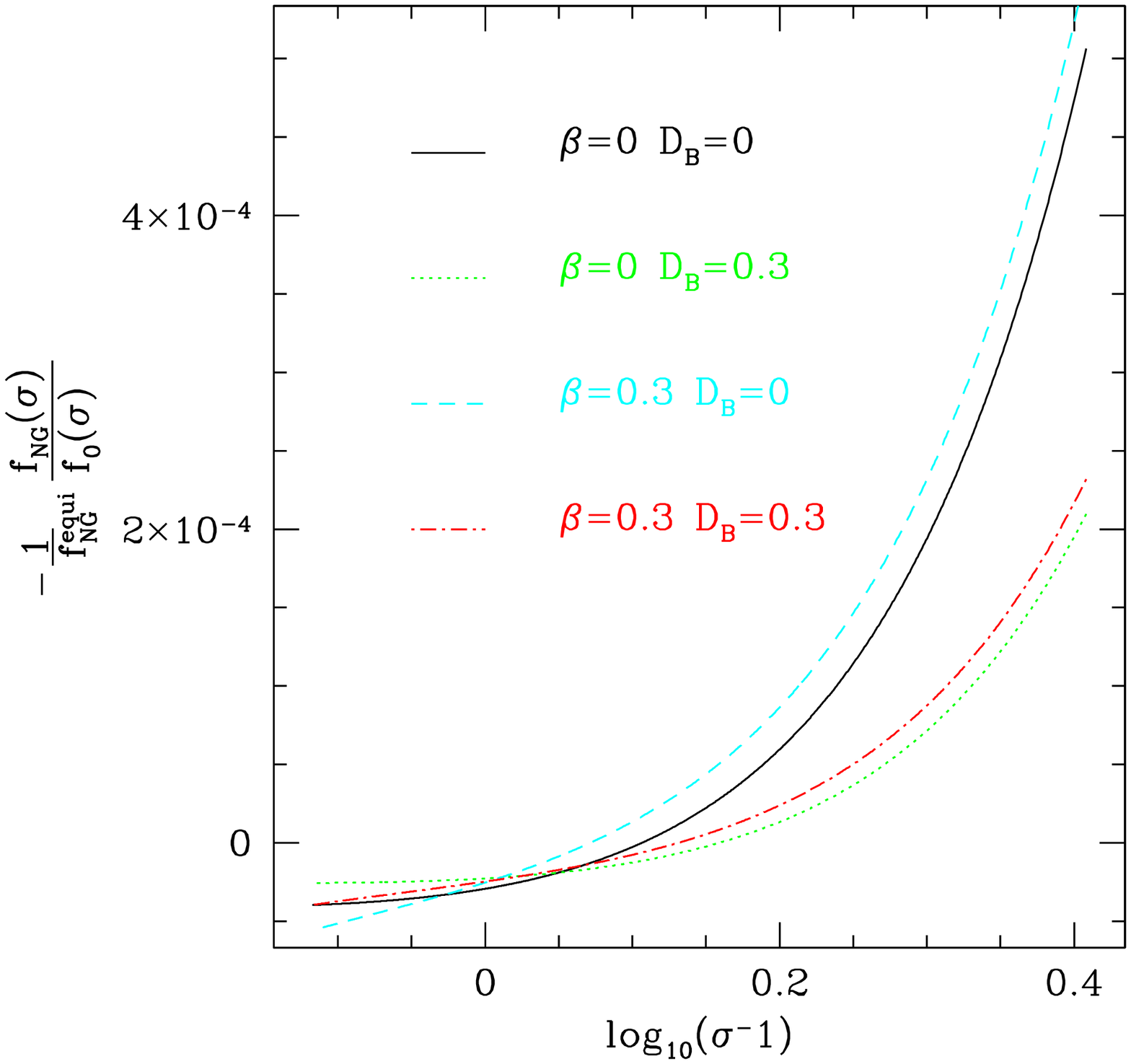}
\end{tabular}
\caption{$f_{NG}(\sigma)/f_0(\sigma)$ as a function of $\log_{10}(\sigma^{-1})$ for local (top panel) and equilateral (bottom panel) non-Gaussianities in units of $f_{NL}$. Different curves correspond to the collapse parameters values of models shown in figure \ref{fig1}}\label{fig5}
\end{figure}

We have inferred the halo mass function in the presence of primordial
non-Gaussianity as an expansion about the
Gaussian Markovian solution by Taylor expanding the path-integral
eq.~(\ref{piepsexp}) to first order in the three-point correlation
function. In this section we derive limits on $f_{NL}$  that bound the validity of such a perturbative
approach by imposing the ratio $|f_{NG}(\sigma)/f_0(\sigma)|<<1$ with $f_{NG}(\sigma)=f_{NG}^{L}(\sigma)+f_{NG}^{NL}(\sigma)$.

In figure~\ref{fig5} we plot $f_{NG}(\sigma)/f_0(\sigma)$ for local (left panel) 
and equilateral (right panel) non-Gaussianities in units of $f_{NL}$ for the barrier model parameters shown in
figure~\ref{fig1}. First, we may notice that in the local case the ratio increases
toward large value of $\log_{10}\sigma^{-1}$ and maximizes in the
high-mass end. Secondly, the overall amplitude of this ratio varies
for different barrier model parameter values. This is an indication
that the non-Gaussian imprint on the halo mass function depends on the
collapse model which we will discuss in more detail in the next
section. A similar trend occurs in the equilateral case, the main
difference result in the low-mass behavior and the value of the ratio
in the high-mass end.

From these curves we infer the ratio $f_{NG}(\sigma)/f_0(\sigma)\sim 0.5$  for 
$|f_{NL}^{\textrm{loc}}|\approx 1000$ in the case of the Diffusive Barrier while for rigid barriers, (i.e. $D_B=0$), we find $|f_{NL}^{\textrm{loc}}|\approx 170$. For the equilateral type the limits are $|f_{NL}^{\textrm{equi}}|\approx 1000$  for the 
rigid barrier and $|f_{NL}^{\textrm{equi}}|\approx 2500$ for the diffusing barriers respectively.


The mass function derived here can be used to compared with N-body simulation and for observational analysis purposes 
to test primordial NG for a large amount of primordial non-Gaussianities and mass range.\footnote{For instance we performed the analysis for $2.10^{13}<M<2.10^{15}h^{-1}M_{\odot}$ at $z=0$.}

\section{Halo Collapse Model and NG Signal}\label{imprints}
\begin{figure}
\centering
\begin{tabular}{cc}
\includegraphics[width=9cm]{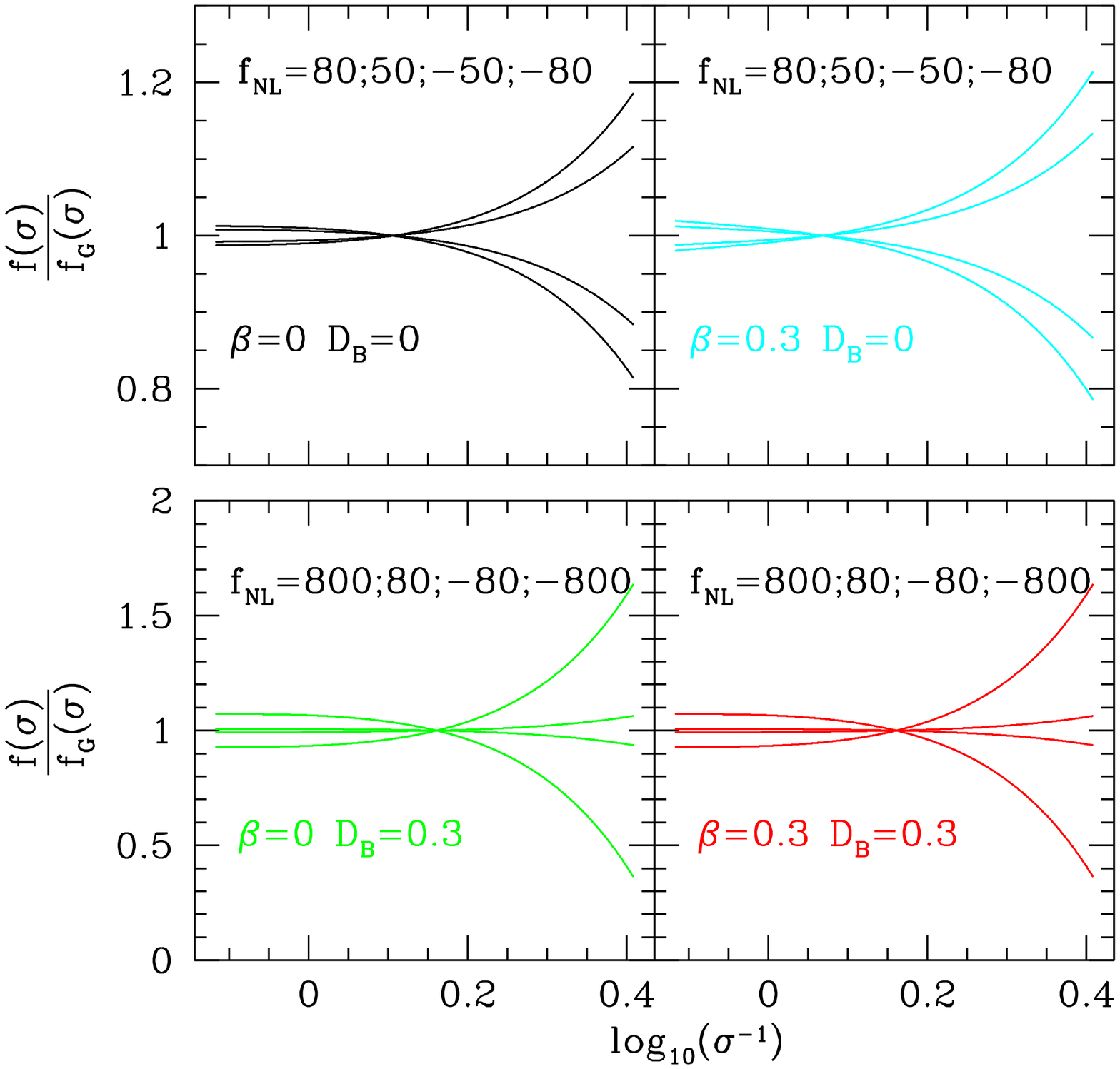} \\
\includegraphics[width=9cm]{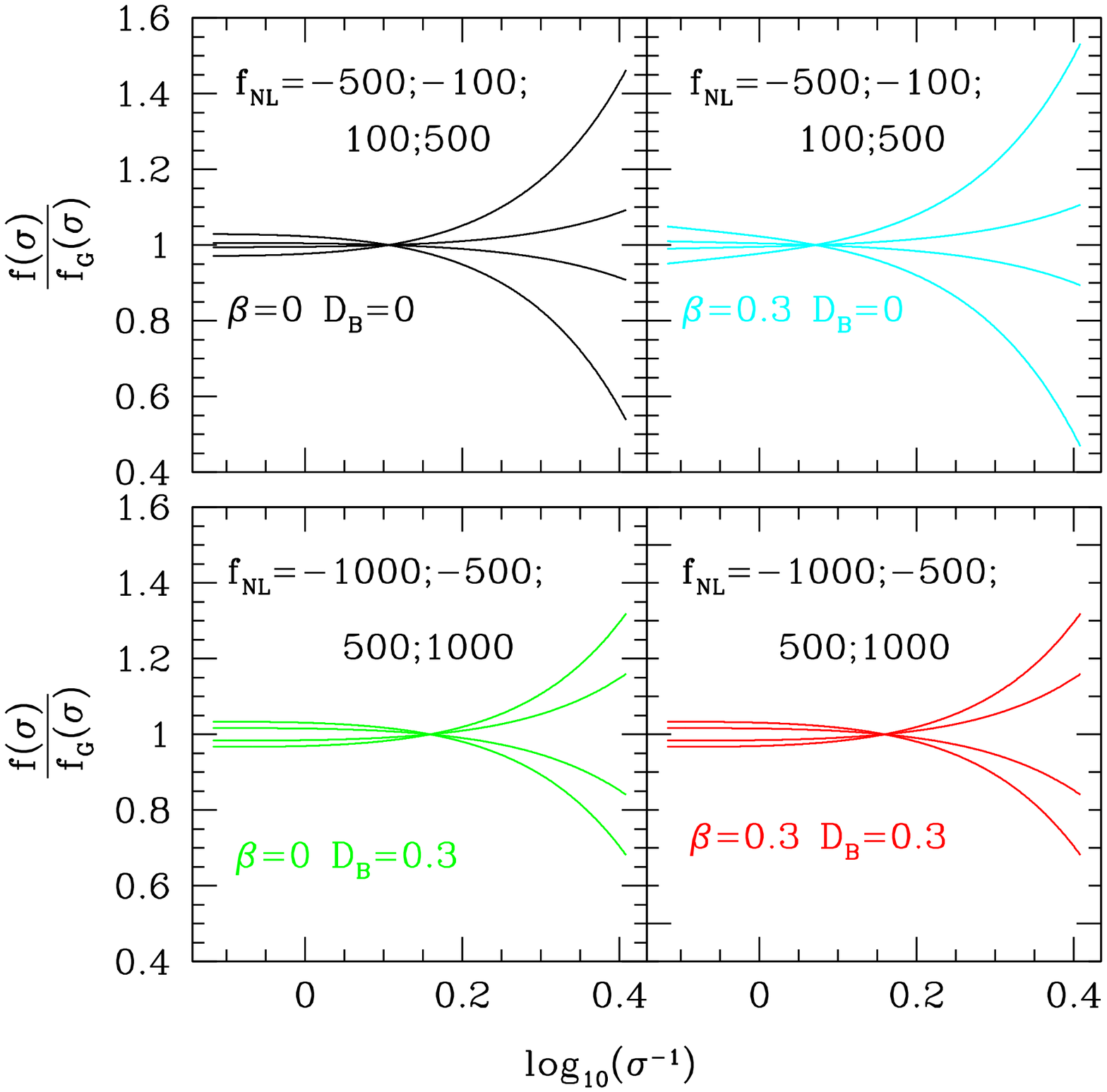}
\end{tabular}
\caption{$\mathcal{R}=f(\sigma)/f_G(\sigma)$ as function of
$\log_{10}(\sigma^{-1})$ for local (top panel) and equilateral (bottom
  panel) non-Gaussianities using different values of the collapse parameters and different non-Gaussian amplitudes.}\label{fig6}
\end{figure}

We now discuss the imprints of primordial non-Gaussianity on 
the halo mass function and the dependence on the halo collapse model
parameters. In order to evaluate the NG features it has become 
standard to consider the ratio $\mathcal{R}=f(\sigma)/f_G(\sigma)$,
which we plot in figure~\ref{fig6} for local NG (top panel) and
equilateral NG (bottom panel) with $f_{NL}=-80,-50,50,80$ in the case of non rigid barriers ($D_B=0$) and $f_{NL}=-800,-80,80,800$ in case of diffusive ones.
We can see that independently of the value of $\beta$ and $D_B$ the deviation from
the Gaussian case maximizes in the high-mass end consistently with
expectations that NG affects primarily the large masses both in the
local and equilateral case. However, we also notice that 
the non-Gaussian signature depends on the barrier model parameter values. 
In particular, by comparing curves with the same value of $f_{NL}$ and
different values of $\beta$ and $D_B$ we can infer two distinctive
trends. On the one hand the
average drift of the barrier enhances the NG imprint, especially at small masses. On the other hand, the barrier diffusion strongly supress 
the overall amplitude of the non-Gaussian signal relative to the
Gaussian one, particularly in the high-mass end. Similar trends occur in the equilateral case. 

As already inferred from the form of the non-Gaussian terms
in section~\ref{ngmass}, this shows that the effects of the
statistics of the primordial matter density field and the non-spherical collapse of halos
are coupled.

Furthermore the nature of the imprints is consistent with the excursion set picture. In
fact, primordial non-Gaussianity manifests as an additional non-Markovian
correlation to the random walk trajectories. The barrier diffusion increases the likelihood
of first-up crossing, while the average drift reduces it and 
as discussed in section~\ref{gaussmass}, the former has the effects of
diminishing the amplitude of the non-Markovian term relative to the
Markovian one, while the latter enhances it. 

Furthermore, the fact that the non-Gaussian signal depends on the halo collapse 
is particularly important from an observational point of view.
In fact, if we don't use the right halo mass function we might put wrong prior on the amount of non-Gaussianity.
It is plausible that such degeneracies may be broken including
measurements of the redshift evolution the halo mass
function. Although, this will require predicting the redshift
and cosmology of the barrier model parameters for non-Gaussian initial
conditions, a study which goes beyond the scope of this paper.

\section{NG Halo Mass Function and N-body Simulations}\label{nbodycomparison}
In recent years, several groups have performed numerical N-body
simulations with non-Gaussian initial conditions to probe the imprints
of primordial non-Gaussianity on the distribution of dark matter
halos \cite{Desjacques09,Grossi09,Pillepich10,Wagner10}. These studies
have provided fitting functions for the non-Gaussian halo mass function
and tested the validity of analytical formula based on the
Press-Schechter formalism. Here, we confront the mass
function derived in the previous sections against the results of 
Pillepich, Porciani and Hahn \cite{Pillepich10}.
These authors have performed a series of high-resolution runs 
with $1024^3$ particles in large volume boxes
resolving halos with mass $2\times 10^{13}<M[\textrm{h}^{-1} M_\odot]<2\times 10^{15}$ at z=0
for local NG initial conditions in the case of WMAP-3yr and WMAP5-yr calibrated cosmologies.
Halo mass functions have been measured using the Friend-of-Friend
algorithm with linking length $b=0.2$ at different redshifts and
fitted with the standard 
parametrization
\begin{equation}\label{mfit}
f_{\textrm{fit}}(\sigma)=\left[D+B\left(\frac{1}{\sigma}\right)^A\right]e^{-\frac{C}{\sigma^2}},
\end{equation}
in doing so the authors have inferred polynomial fits which parameterize the
dependence on $f_{NL}$ and $z$ of the coefficients A, B, C and D. Here, we
limit our theoretical model comparison with the results at $z=0$.
In what follow we use the parameters of Table 5 of  \cite{Pillepich10} for which the accuraty is of order $5\%$ on the range $-0.2<\ln(\sigma^{-1})<0.8$

Before discussing the results of this comparison let us make a few
remarks. The excursion set mass function for the diffusive
barrier model depends on two input parameters which specifies the
statistical properties of the stochastic barrier. Other quantities 
such as the spherical collapse threshold $\delta_c$
and the amplitude of the filter-correction $\kappa$ 
are predicted for a given cosmological model. For instance in $\Lambda$CDM
WMAP5-yr cosmology $\delta_c=1.673$ at $z=0$ and $\kappa=0.475$.

In principle, the values of $\beta$ and $D_B$ can be predicted 
for a given cosmology. As already stressed in section~\ref{stocmod} this
requires estimating the properties of the stochastic barrier by
numerically solving the ellipsoidal collapse model equations. As shown by Doroshkevich
\cite{Doroshkevich1970}, the randomness of the initial density field
causes the parameters describing a homogeneous ellipsoid to be random
variable themselves with a characteristic probability distribution. By
generalizing this to non-Gaussian initial conditions and solving 
the ellipsoidal collapse model equations one can infer
the probability distribution of the collapse threshold as function of
the variance of the linear density field. Then, the
stochastic barrier model parameters can be derived by matching their
value to the moments of the probability distribution. Such an
inference will provide a complete prediction of the cosmology, redshift
and non-Gaussian dependence of $\beta$ and $D_B$ which at moment
is missing. However, testing such dependences against numerically inferred N-body halo
mass functions will first require addressing the effect of the second order cloud-in-cloud
problem, which as we have mentioned in Section \ref{stocmod} manifests as an additional scatter on the halo collapse
threshold.
\begin{figure}[h]
\centering
\includegraphics[width=10cm]{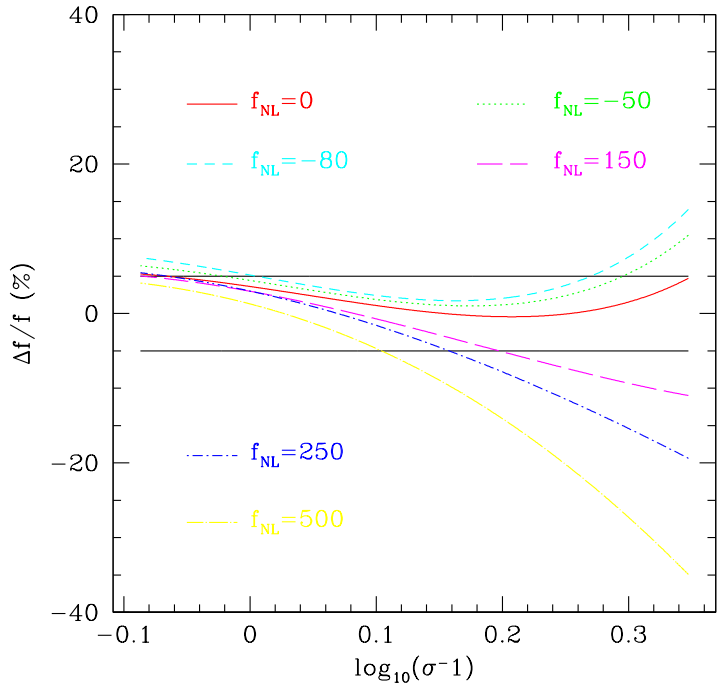}
\caption{Relative difference of the NG mass function with respect to
  Pillepich et al. fitting formula for the Gaussian calibrated barrier
  model parameters. The discrepancy remains within the numerical
  accuracy of the N-body results (solid parallel lines) for $-10\lesssim f_{NL}\lesssim 30$.}\label{fig7}
\end{figure}

In the absence of theoretical prediction for the values of $\beta$ and
$D_B$, the diffusive drifting barrier model remains an effective
description of the non-spherical collapse of halos. Consequently, 
the model parameters can be determined only
through comparison with N-body simulation results. For instance, 
in \cite{PSCIA2011a,PSCIA2011b} we have confronted the Gaussian mass
function N-body simulation data by Tinker
et al. \cite{Tinker2008} obtained using the Spherical Overdensity (SO)
algorithm. This has allowed us to infer the best fit values 
$\beta_G^{SO}=0.057$ and $D_B^{SO}=0.294$ and quite remarkably found that for such
parameter values the mass function 
reproduces the data to numerical accuracy ($\sim 5\%$). However, we
cannot assume these values to confront with the results from the
Pillepich et al. simulations even for $f_{NL}=0$. This is because there is roughly 
a $\sim 10\%$ scatter between SO and FoF mass functions due to
systematic differences of the halo detection algorithms. 
Therefore we have calibrated the value of $\beta$ and $D_B$
against the Gaussian mass function from \cite{Pillepich10}.

We find
$\beta^G=0.0529$ and $D_B^G=0.328$.
Assuming such values we plot in figure~\ref{fig7} the relative
difference between eq.~(\ref{ftot}) and eq.~(\ref{mfit}) for different values of $f_{NL}$.
We can see that for $f_{NL}=0$ the differences are well within the
numerical uncertainties $\approx5\%$, which is consistent with the
results of the comparison to Tinker et al. \cite{Tinker2008} presented
in \cite{PSCIA2011a,PSCIA2011b}. As it can be seen in
figure~\ref{fig7}, for $f_{NL}^{\rm loc}\ne 0$ the relative difference
remains within the numerical errors only in the range $-50\lesssim f_{NL}\lesssim 150$, while
for larger values the mass function evaluated using the Gaussian
calibrated parameters largely deviates from the numerical
fit.

This simply implies that for $f_{NL}$ outside this range
the non-spherical collapse of halos strongly differ from the Gaussian
case and consequently we can expect $\beta$ and $D_B$ to vary with
$f_{NL}$. 

We can see this more clearly by inferring the best fit values of $\beta$ and
$D_B$ for different values of $f_{NL}$ in the range $-80<f_{NL}<750$
for which we compare equation~(\ref{ftot}) against eq.~(\ref{mfit}). We may notice that
$\beta^{\rm fit}$ increases for increasing values of $f_{NL}>0$,
as well as $D_B^{\rm fit}$.

\begin{figure}[h]
\centering
\includegraphics[width=10cm]{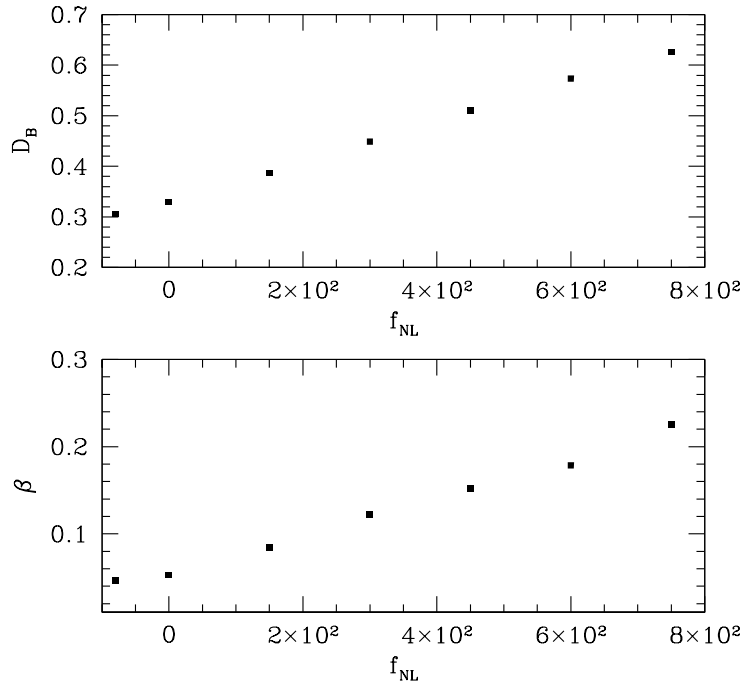}
\caption{Best fit values of $\beta$ (top panel) and $D_B$ (bottom
 panel) relative to the Gaussian
  case as function of $f_{NL}$.}\label{fig8}
\end{figure}

In particular, for large positive values of $f_{NL}$ the average drift 
is about $4$ times larger than in the Gaussian case. This indicates
that for large primordial non-Gaussianity the collapse of dark matter
halos is highly non-spherical. Furthermore,  since the drift coefficient increases for increasing values of $f_{NL}$, halos with higher mass will be privileged relative to the smaller ones.

Similarly, the diffusion increases with respect to the Gaussian case for increasing values of $f_{NL}$. It is worth reminding that the barrier has a Gaussian distribution of mean value $\bar{B}=\delta_c+\beta S$ and variance $D_{B}$. Therefore, for positive value of $f_{NL}$, the non-Gaussian nature of the initial density field tends to erase the imprint of the mean ellipsoidal collapse value which carries the  signature of the gravitational dynamics.

It would be interesting to perform a study similar to that of
Robertson et al. \cite{Robertson2009} to infer the distribution of the linear collapse threshold 
of halos detected in non-Gaussian simulations, which can provide a direct
confirmation of the trend of $\beta$ and $D_B$ as function of $f_{NL}$
obtained here.\\
\\

\begin{figure}[h]
\centering
\includegraphics[width=15cm]{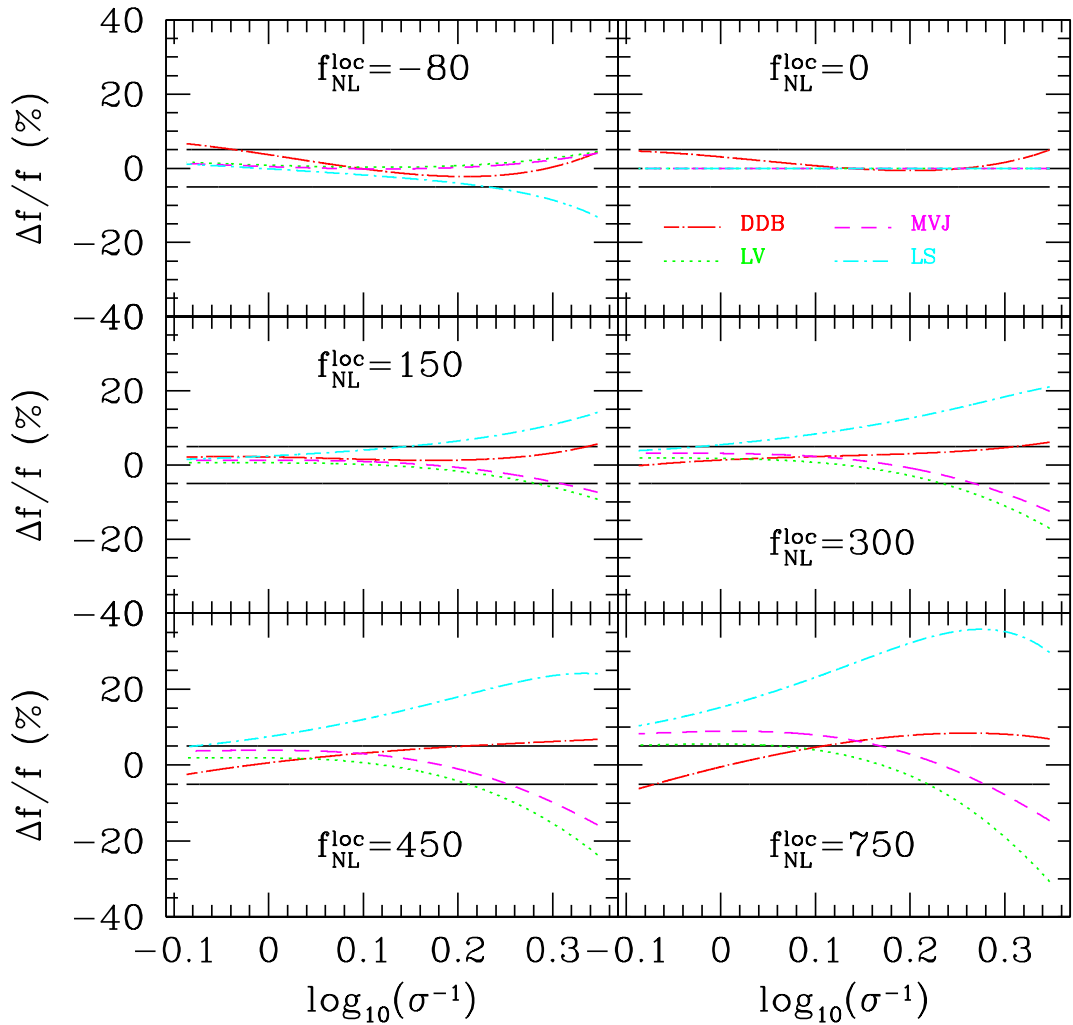}
\caption{Relative difference with respect to the Pillepich et
  al. fitting formulae for different values of $f_{NL}^{\rm loc}$ of non-Gaussian halo mass functions for the DDB
  case with best fitting barrier parameter values (red long-dash dot line), Loverde et
  al. \cite{Loverdeetal2008} (green dot line),
  Matarrese-Jimenez-Verde \cite{Matarrese2000} (magenta short-dash
  line) and Lam-Sheth \cite{LamSheth2009} (cyan short-dash dot
line) respectively.}\label{fig9}
\end{figure}
In figure~\ref{fig9} we show the relative difference with respect to Pillepich et al. fitting formula of 
the NG mass function computed with the best fit barrier parameters
(red long-dash dot line)
for the values of $f_{NL}^{loc}$ used in figure~\ref{fig8}. We can see that the
differences remain within the numerical accuracy of the N-body results.
In addition, we show the relative difference for the Loverde et
al. \cite{Loverdeetal2008} formula (LV, green dot line),
Matarrese-Verde-Jimmenez \cite{Matarrese2000} (MVJ, magenta short-dash
line) and Lam-Sheth \cite{LamSheth2009} (LS, cyan short-dash dot
line). Since the Gaussian limit of these different formulae is given
by the Press-Schechter mass function, for a comparison with the
N-body simulation results we have renormalized them to the Gaussian fitting formula by Pillepich et
al. \cite{Pillepich10}. We find these formulae to be close to the numerical accuracy
of the N-body results for $-80<f_{NL}^{loc}<150$. As we infered in from ~\ref{fig6} this correspond to the range where the non-Gaussian dependence
of the non-spherical collapse is negligible. In the light of this analysis we find that our halo mass function explains the so called ''fudge" factor which is used to fit the non-Gaussian simulation. Moreover outside the range $-50<f_{NL}^{loc}<150$
the non-spherical nature of the halo collapse must be
taken into account when testing primordial non-Gaussianity with 
halo mass function measurements. 


\section{Conclusion}\label{conclu}
In the upcoming years several observational program will provide
precise measurements of the number counts and spatial distribution of
galaxy clusters. These measurements will provide new insights on the
mass distribution of dark matter halos and consequently on the
underlying cosmology and the statistics of the primordial density field. 

Here, we have performed an {\it ab initio} calculation of the NG halo mass function using the
path-integral formulation of the excursion set formalism. We have
considered a stochastic model of the halo collapse threshold 
which captures the main features of the non-spherical collapse
of halos. We have specifically focused on primordial NG due to a non-vanishing skewness of
the initial density field and inferred the mass function through a
perturbative expansion of the path-integral. The computation can be extended to
higher-order correlations and we have showed that the effect of a
non-vanishing kurtosis can be easily obtained at leading order. 

Contrary to standard approaches based on NG extension of the
Press-Schechter calculation, we find that the effects of the non-spherical nature of
the collapse of halos on the mass function directly couples to the
signature of primordial non-Gaussianitiy. As such, these effects
cannot be simply reabsorbed in a Gaussian normalization term. 
We have studied the case for local and equilateral non-Gaussianities,
though the formulae provided here can be applied to any NG shape function.

We have compared the inferred mass function with NG local
N-body simulations results and found a remarkable agreement. 
Using the barrier model parameters calibrated on the Gaussian N-body
mass function we find that for large non-Gaussianities ($f_{NL}<-50$ and
$f_{NL}>150$) the non-spherical collapse largely deviates from the
Gaussian case and its effect cannot be simply reabsorbed in a Gaussian
normalization as in the case of PS derived mass functions. 
These effects can, therefore, be important for cluster count studies
since the non-Gaussian dependence of the non-spherical collapse of
halos changes the imprint of primordial non-Gaussianities on the mass
function. For instance in \cite{LeeJ} the author has shown that the
abundance of clusters which do not belong to the super clusters, detected in Gaussian N-body
simulations, is well described by the Gaussian mass function derived in
\cite{PSCIA2011a,PSCIA2011b} for $D_{B}=0$ and a non-vanishing $\beta$. In light of this result we can expect that the number counts of this type of clusters may be more sensitive to the signature of primordial NG since the
imprint has a larger amplitude compared to field clusters with $D_B\ne 0$ and $\beta=0$. 

These results suggest several directions of future investigation. On
the theoretical side it appears evident that a theoretical prediction
of $\beta$ and $D_B$ based on a detailed study of the
ellipsoidal collapse model is needed. This will allow us to 
predict the cosmology and redshift dependence as well as their relation
with respect to statistics of the primordial density field. Thus, a
barrier model parameter inference using mass function measurements may
provide a way of distinguishing between different cosmological
scenarios assuming a proper treatment of the second order cloud-in-cloud
problem. On the numerical side it will be interesting to confront
these model predictions with the linearly extrapolated collapse
threshold of halos detected in N-body simulations for different
cosmological scenarios.

\acknowledgments 
We thank Ravi Sheth and Tommaso Giannantonio for useful discussions and comments. Research for this work is supported in part by a grant of the University Paris Diderot.

\end{document}